\documentclass[prd,preprintnumbers,nofootinbib,superscriptaddress,amsmath,english]{revtex4}
\usepackage{epsfig,amsmath} 
\usepackage{amssymb}
    \usepackage{graphics}
     \usepackage{bm}
      \usepackage{pifont}
      \usepackage{color}
\usepackage{yfonts}

\def\lsim{\mathrel{\rlap{\lower4pt\hbox{\hskip1pt$\sim$}}
    \raise1pt\hbox{$<$}}}                
\def\gsim{\mathrel{\rlap{\lower4pt\hbox{\hskip1pt$\sim$}}
    \raise1pt\hbox{$>$}}}                

\newcommand{\balign}{\begin{align}}
\newcommand{\ealign}{\end{align}}

\newcommand{\BFKL}{Kuraev:1976ge,Kuraev:1977fs,Balitsky:1978ic}

\newcommand{\DGLAP}{Gribov:1972ri,Lipatov:1974qm,Altarelli:1977zs,Dokshitzer:1977sg}

\def\tmdlib{{\sc TMDlib}}
\def\tmdplotter{{\sc TMDplotter}}
\def\cascade{{\sc Cascade}}
\def\pythia{{\sc Pythia}}

\begin{document}

\begin{flushright}
DESY 15-111\\
NIKHEF 2015-023\\ 
RAL-P-2015-006 
\end{flushright}
\vspace*{-3.4cm}
\title{Transverse momentum dependent (TMD) 
parton distribution  functions:\\ 
 status and prospects}

\author{R.~Angeles-Martinez}
\affiliation{School of Physics and  Astronomy, University of Manchester, UK}

\author{A.~Bacchetta}
\affiliation{INFN Sezione di Pavia and 
Dipartimento di Fisica,  Universit{\` a} di Pavia, Italy}

\author{I.I.~Balitsky}
\affiliation{Physics Department, ODU and Theory Group, JLab, USA}

\author{D.~Boer}
\email{d.boer@rug.nl}
\affiliation{Van Swinderen Institute, University   of Groningen, The Netherlands}

\author{M.~Boglione}
\affiliation{Dipartimento di Fisica, Universit{\` a} 
 di Torino and INFN, Torino, Italy}

\author{R.~Boussarie}
\affiliation{LPT, Universit{\' e}  Paris-Sud, CNRS, Orsay, France}

\author{F.A.~Ceccopieri}
\affiliation{FPA,  Universit{\' e} de Liege,  Belgium}

\author{I.O.~Cherednikov}
\email{igor.cherednikov@uantwerpen.be}
\affiliation{Universiteit Antwerpen, Belgium}

\author{P.~Connor}
\affiliation{DESY, Germany}

\author{M.G.~Echevarria}
\affiliation{Nikhef Theory Group and VU University, Amsterdam, the Netherlands}

\author{G.~Ferrera}
\affiliation{Dipartimento di Fisica, Universit{\` a} 
 di Milano and INFN, Milano, Italy}

\author{J.~Grados~Luyando}
\affiliation{DESY, Germany}

\author{F.~Hautmann}
\email{f.hautmann1@physics.ox.ac.uk}
\affiliation{RAL,   University of Oxford and   University of Southampton,    UK}

\author{H.~Jung}
\email{hannes.jung@desy.de}
\affiliation{Universiteit Antwerpen, Belgium}
\affiliation{DESY, Germany}  

\author{T.~Kasemets}
\affiliation{Nikhef Theory Group and VU University, Amsterdam, the Netherlands}

\author{K.~Kutak}
\affiliation{Instytut Fizyki Jadrowej Polskiej Akademii Nauk, Krakow, Poland}

\author{J.P.~Lansberg}
\affiliation{IPNO, Universit{\' e}  Paris-Sud, CNRS/IN2P3, Orsay, France}

\author{A.~Lelek}
\affiliation{DESY, Germany}

\author{G.~Lykasov}
\affiliation{JINR Dubna, Russia}

\author{J.D.~Madrigal~Martinez}
\affiliation{Institut de Physique 
Th{\' e}orique, CEA Saclay, 
CNRS, Gif-sur-Yvette, France}

\author{P.J.~Mulders}
\email{p.j.g.mulders@vu.nl}
\affiliation{Nikhef Theory Group and VU University, Amsterdam, the Netherlands}

\author{E.R.~Nocera}
\affiliation{Dipartimento di Fisica, Universit{\` a} 
 di Genova and INFN, Genova, Italy}

\author{E.~Petreska}
\affiliation{Centre de Physique 
Th{\' e}orique, {\' E}cole Polytechnique, Palaiseau, France}  
\affiliation{Departamento de 
F\'isica de Particulas / IGFAE, 
Univ. de Santiago 
de Compostela, Spain}

\author{C.~Pisano}
\affiliation{Universiteit Antwerpen, Belgium}

\author{R.~Pla{\v{c}}akyt{\.{e}}}
\affiliation{DESY, Germany}

\author{V.~Radescu}
\affiliation{DESY, Germany}

\author{M.~Radici}
\affiliation{INFN Sezione di Pavia and Dipartimento di Fisica,  Universit{\` a} di  
Pavia, Italy}   

\author{G.~Schnell}
\affiliation{University of the Basque Country UPV/EHU and IKERBASQUE, Bilbao, Spain}

\author{I.~Scimemi}
\affiliation{Departamento de F\'isica Te\'orica II, Universidad Complutense de Madrid, Spain}

\author{A.~Signori}
\email{asignori@nikhef.nl}
\affiliation{Nikhef Theory Group and VU University, Amsterdam, the Netherlands}

\author{L.~Szymanowski}
\affiliation{National Centre for Nuclear Research (NCBJ), Warsaw, Poland}

\author{S.~Taheri~Monfared}
\affiliation{School of Particles and Accelerators, Institute for Research in Fundamental Sciences (IPM), Tehran, Iran}

\author{F.F.~Van~der~Veken}
\affiliation{Universiteit Antwerpen, Belgium}

\author{H.J.~van Haevermaet}
\affiliation{Universiteit Antwerpen, Belgium}

\author{P.~Van Mechelen}
\email{pierre.vanmechelen@uantwerpen.be}
\affiliation{Universiteit Antwerpen, Belgium}

\author{A.A.~Vladimirov}
\affiliation{Department of Astronomy and Theoretical Physics, Lund University, Sweden}

\author{S.~Wallon}
\affiliation{LPT, Universit{\' e}  Paris-Sud, CNRS, Orsay, France}
\affiliation{Universit{\' e}   Paris 06, Facult{\' e}  de Physique, Paris, France}


\vspace {20mm}
\date{\today}

\vspace {20mm}
\begin{abstract}
We provide a concise overview 
on 
 transverse momentum dependent (TMD) 
parton distribution  functions, 
their application to topical issues 
in high-energy physics 
phenomenology, and their 
theoretical connections with QCD resummation, evolution 
and factorization theorems.  We 
illustrate the use of TMDs via 
examples of multi-scale problems 
in hadronic collisions. These 
 include transverse momentum 
$q_T$ spectra of Higgs and 
vector  bosons for low $q_T$, and  
azimuthal correlations in the production of  
multiple jets associated with heavy bosons at large jet 
masses.  We discuss computational 
tools  for TMDs,  and present an application of  
a new tool, \tmdlib, to parton density fits and parameterizations.  
\end{abstract}

\maketitle

\section{Introduction}

Experimental information on 
``3-dimensional imaging" of 
hadrons,  encoded in 
unintegrated,  transverse 
momentum dependent (TMD)   
parton density and parton decay  
functions, comes at present from 
two main sets of  experimental 
data: deep inelastic scattering 
(DIS) at 
high energy, and  low-$q_T$ 
Drell-Yan (DY) and semi-inclusive 
DIS (polarized and unpolarized). 
In each of these two cases, 
 QCD factorization theorems allow 
one to relate physical, observable  
cross sections 
to TMD parton distributions via 
perturbatively calculable kernels. 
These theorems 
provide the theoretical 
basis for determining  
TMD distributions from 
experimental measurements. 
They are also essential to 
formulate and apply  methods
of  perturbative resummation 
at all orders in the QCD coupling 
to   a large variety of observables 
in high-energy hadronic collisions.  
Examples include  processes both 
at the Large Hadron Collider (LHC)  
and at fixed-target experiments.  

This article is based on workshops 
devoted to these topics held 
at the University of Antwerp in  
2014~\footnote{Workshops on ``Resummation, Evolution, Factorization" (REF 2014), Antwerp, 23-25 June 2014 and 8-11 December 2014.}     
and provides a 
 concise status report of this field. 
The  purpose of this 
article is to give a 
non-technical introduction  to the 
motivations for experimental and 
theoretical studies of TMDs; 
to illustrate this with specific 
examples of application of TMDs 
to topical issues in high-energy 
physics  phenomenology;   
to point to  future  
 directions of development. 

In particular, we examine       implications of 
  two sets of QCD factorization theorems  based on TMD parton 
distribution functions: 
 low-$q_T$ factorization for heavy particle spectra 
(including vector bosons, Higgs bosons, heavy flavors) 
and  high-energy factorization. 
We  focus  on 
production processes in hadronic collisions in two limits:  
i)  $q_T \to 0 $ for fixed invariant mass, and 
ii) $\sqrt{s} \to \infty$ for  fixed momentum transfer. 
We illustrate this with  examples 
on   transverse momentum spectra 
and angular correlations  at the LHC 
 for Drell-Yan and    
  Higgs boson production and 
associated multi-jets.  
We survey computational tools  
which  are 
being developed to treat the physics 
of TMDs. We present in particular 
an application  of a new 
tool, \tmdlib,  to  TMD 
parton densities  based on  
fits and parameterizations 
including QCD evolution. 

The paper   is organized as follows. 
In Sec.~II we motivate the use of  TMDs.   In    Sec.~III  we 
 discuss  their role 
in the physics of large  transverse 
momenta. Sec.~IV summarizes  experimental prospects and  theory 
developments. Secs.~V and VI illustrate the status of   fits and 
parameterizations for TMD parton distributions 
and of TMD Monte Carlo tools.  Final remarks are 
given in Sec.~VII.

\section{Why TMDs}

\begin{figure}[htbp]
\begin{center}
\includegraphics[scale=0.5]{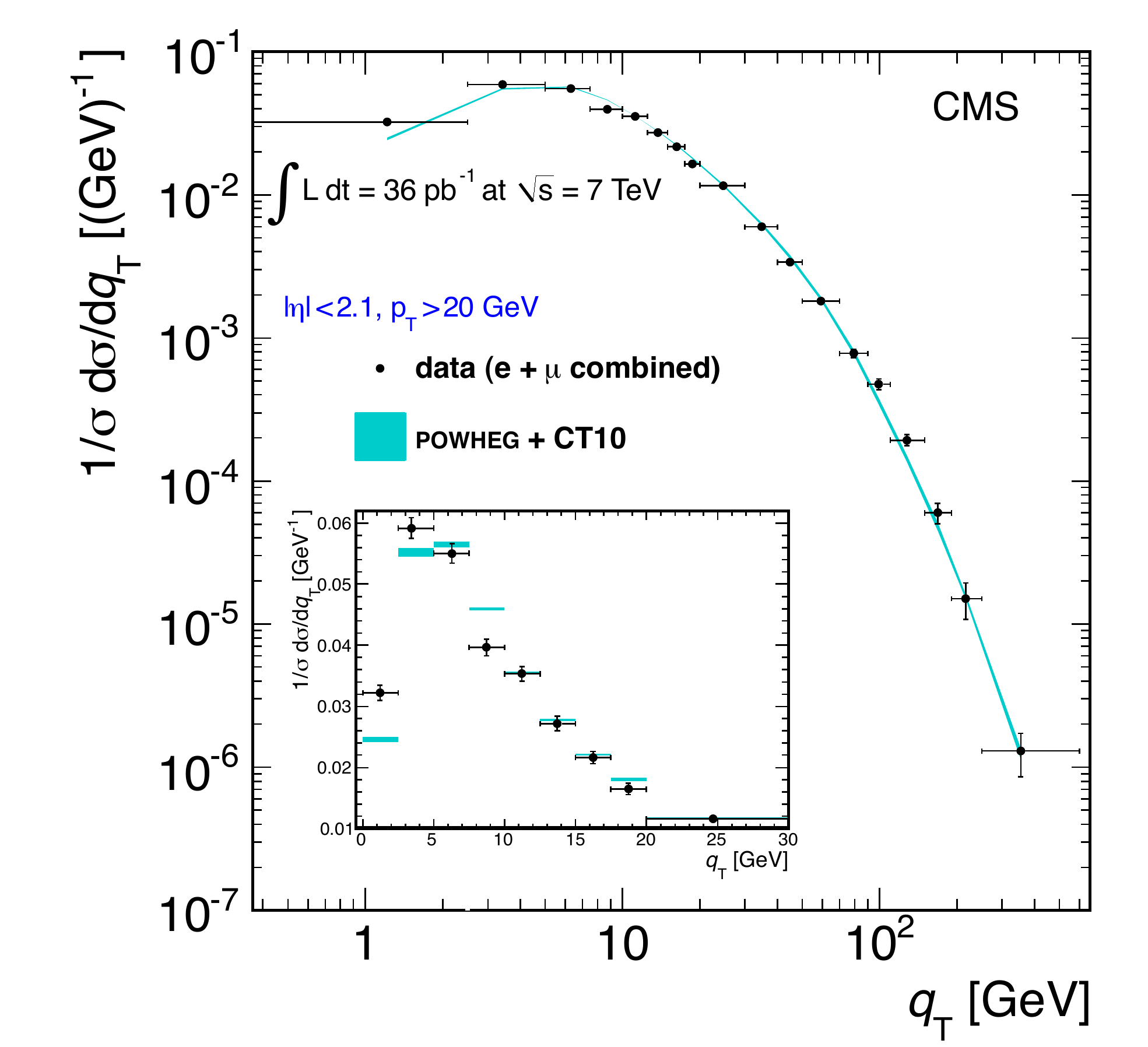}
\caption{\it  The 
$Z$-boson transverse 
momentum $q_T$ spectrum in 
$pp$ collisions 
at the LHC~\protect\cite{Chatrchyan:2011wt}.}
\label{fig:REFfigZ}
\end{center}
\end{figure}

Transverse momentum dependent    parton distributions 
  encode nonperturbative information on hadron structure, 
including transverse momentum and polarization 
degrees of freedom,   
which is  essential  in the context of QCD factorization theorems 
for   multi-scale, non-inclusive   collider  observables. 
A classic example is given by 
Drell-Yan hadroproduction 
of electroweak gauge bosons. 
Fig.~\ref{fig:REFfigZ}~\cite{Chatrchyan:2011wt} 
shows the differential cross section 
for $Z$-boson production  
 in $pp$ collision at the LHC as 
a function of 
the $Z$-boson transverse 
momentum $q_T$, in the lepton pair's     invariant mass 
range 60 GeV $ < M < $ 120 GeV. 
In the spectrum of Fig.~\ref{fig:REFfigZ} 
we  distinguish the  
high-$q_T$ region,  
the peak region, and 
the low-$q_T$ region.

In the high-$q_T$ region
 the cross section 
 is expected to be well represented  
by an evaluation of the 
partonic $Z$-boson cross section 
to finite order in 
QCD perturbation theory 
(leading-order (LO),  
next-to-leading-order (NLO), and so 
forth),     combined with factorization 
in terms of ordinary (collinear)  
parton distribution functions (pdfs).  
On the other hand, if this 
theoretical framework 
 is 
applied to the region of decreasing 
$q_T$ it will not be  
able to describe the 
approach to the 
peak region in Fig.~\ref{fig:REFfigZ}  
 ($q_T \approx 
{\cal O } (10  $ GeV$)$)     
nor  the turn-over  region 
($q_T \approx {\cal O } (1  $~GeV$)$). 
Rather, the cross section predicted 
from any finite order of perturbation   theory, 
convoluted with ordinary 
parton distributions,  
will diverge  as $q_T$ decreases. 
The reason for this is that 
the physical behavior of the 
$Z$-boson spectrum near the peak 
 region and below~\cite{Dokshitzer:1978yd,Parisi:1979se} 
is controlled by multi-parton QCD 
radiation, which is not  well approximated 
by truncating the QCD 
perturbation series to any 
fixed order but rather requires methods 
to resum arbitrarily many 
parton emissions, viz.,    
scattering amplitudes with  
an infinite number of real and 
virtual  insertions of soft 
gluons.  

This can be 
accomplished in a systematic manner 
via a generalized form   of 
QCD factorization~\cite{Collins:1984kg,Collins:1982wa,Collins:2011zzd} which now 
involves quark  distribution functions 
that, unlike the ordinary ones, 
explicitly  depend  on 
transverse momentum and 
polarization (TMD pdfs). 
 Such TMD pdfs obey  
evolution equations~\cite{Collins:1981uw,Collins:1981va,Collins:2011zzd} which generalize
the ordinary renormalization-group 
evolution equations of collinear pdfs. 
These evolution equations, once 
combined with the TMD factorization 
of the physical cross section, 
allow one to resum 
logarithmically enhanced contributions 
in the ratio $M / q_T$ 
to the perturbation series expansions 
  for the  physical observables 
to  all higher orders in the   QCD coupling.  
It is only after this  generalized 
factorization   analysis ---  going 
beyond the collinear 
factorization  --- is 
carried through 
that the physical behavior of the 
$Z$ boson spectrum observed in 
Fig.~\ref{fig:REFfigZ}  
can be predicted.

\begin{figure}[htbp]
\begin{center}
\includegraphics[scale=0.65]{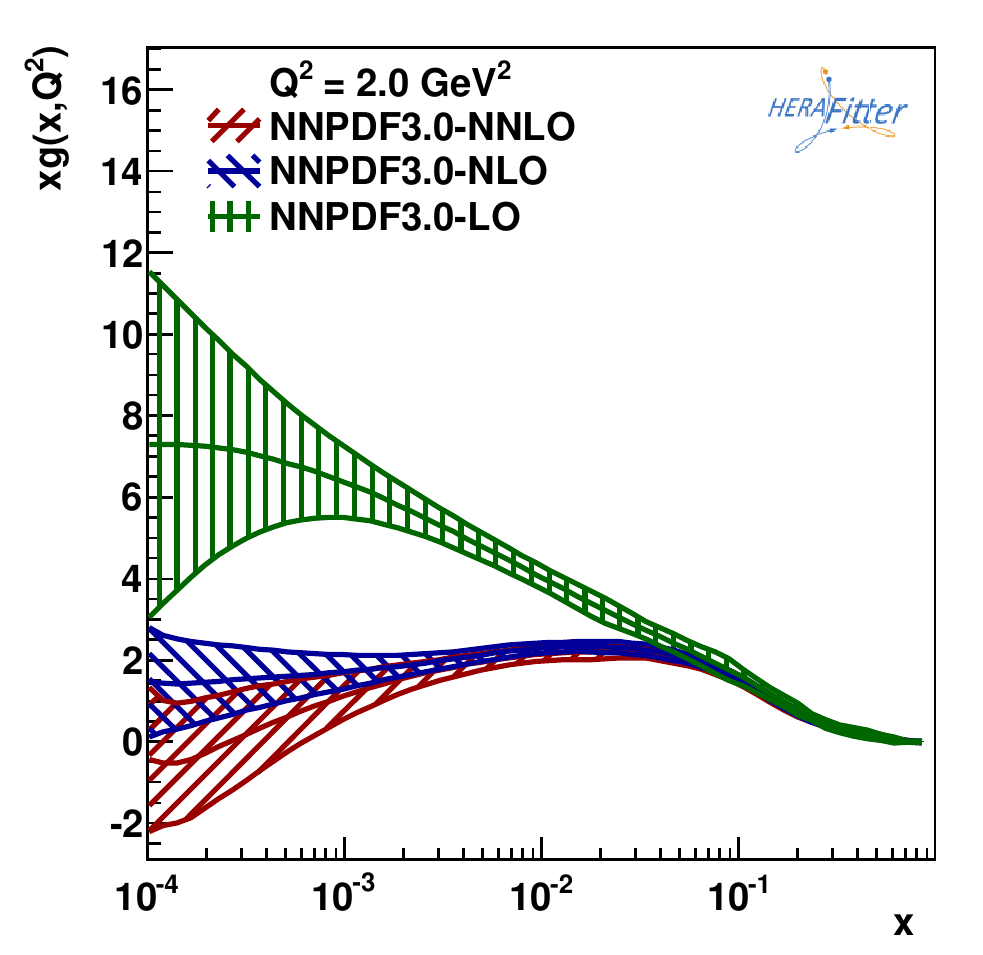}
\includegraphics[scale=0.65]{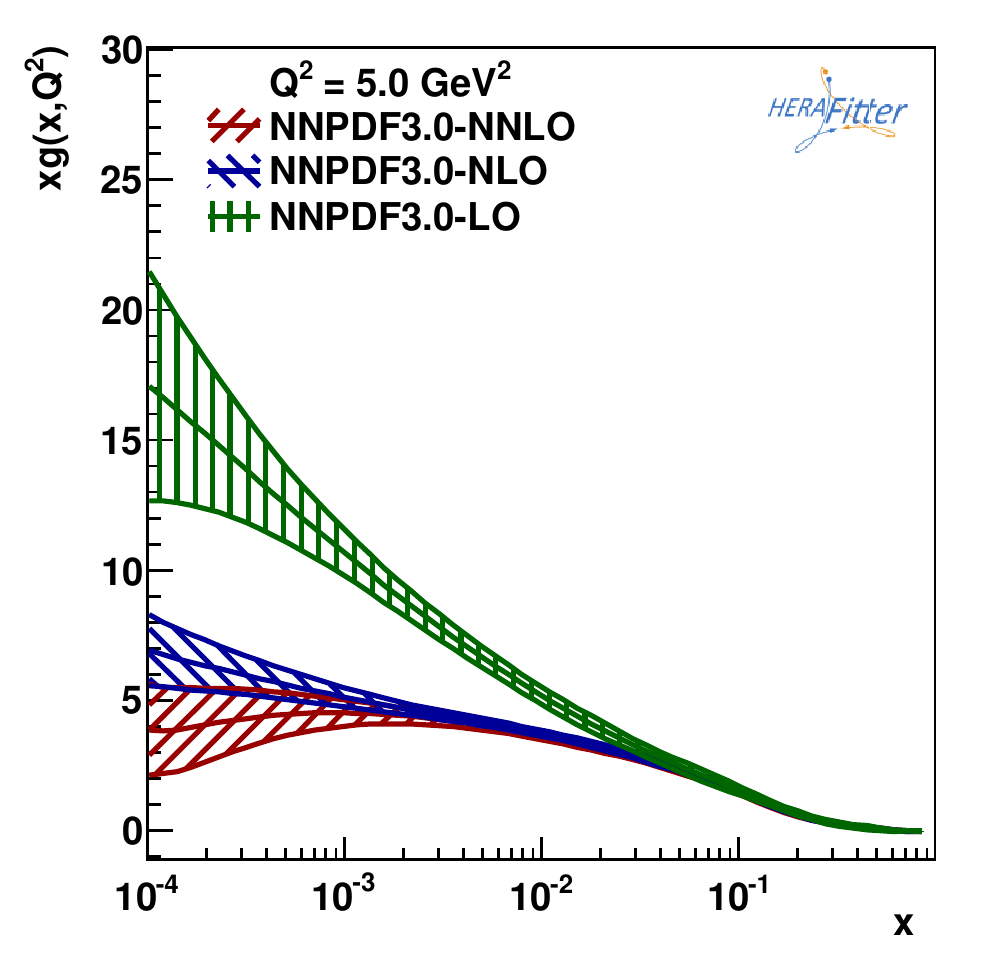}
\includegraphics[scale=0.65]{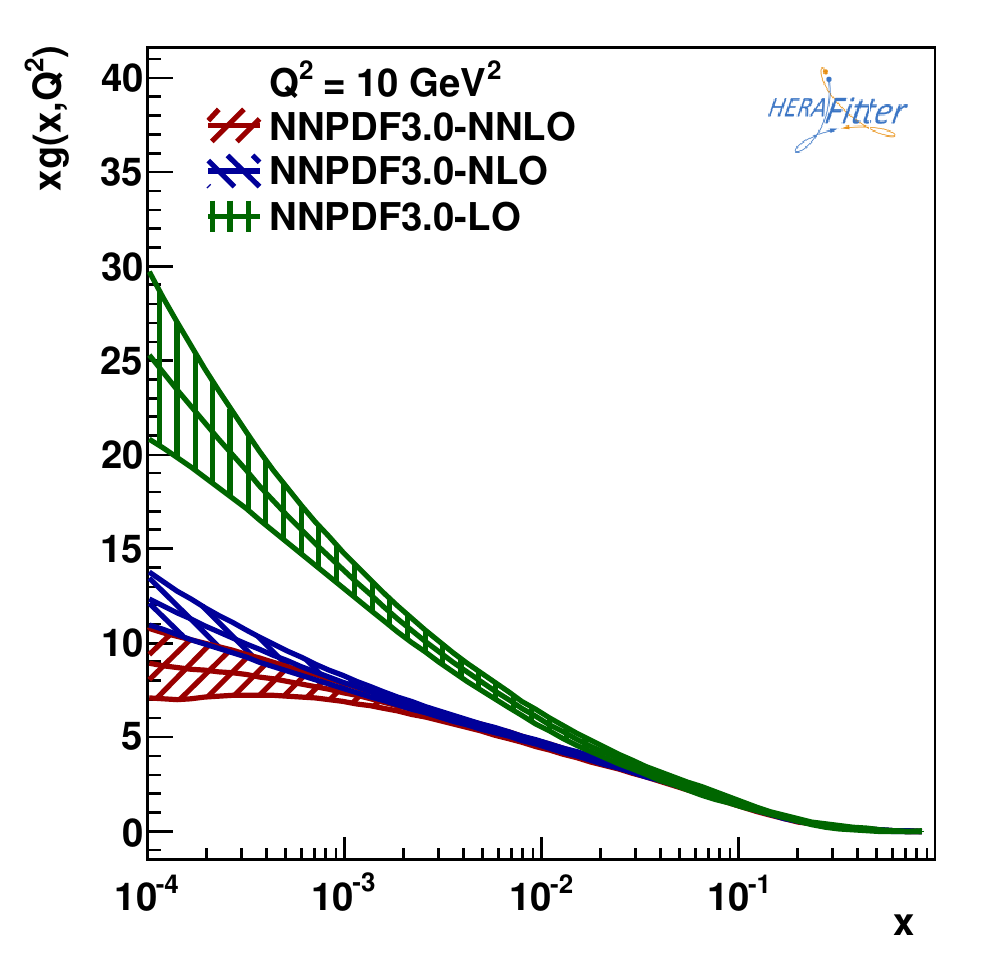}
\includegraphics[scale=0.65]{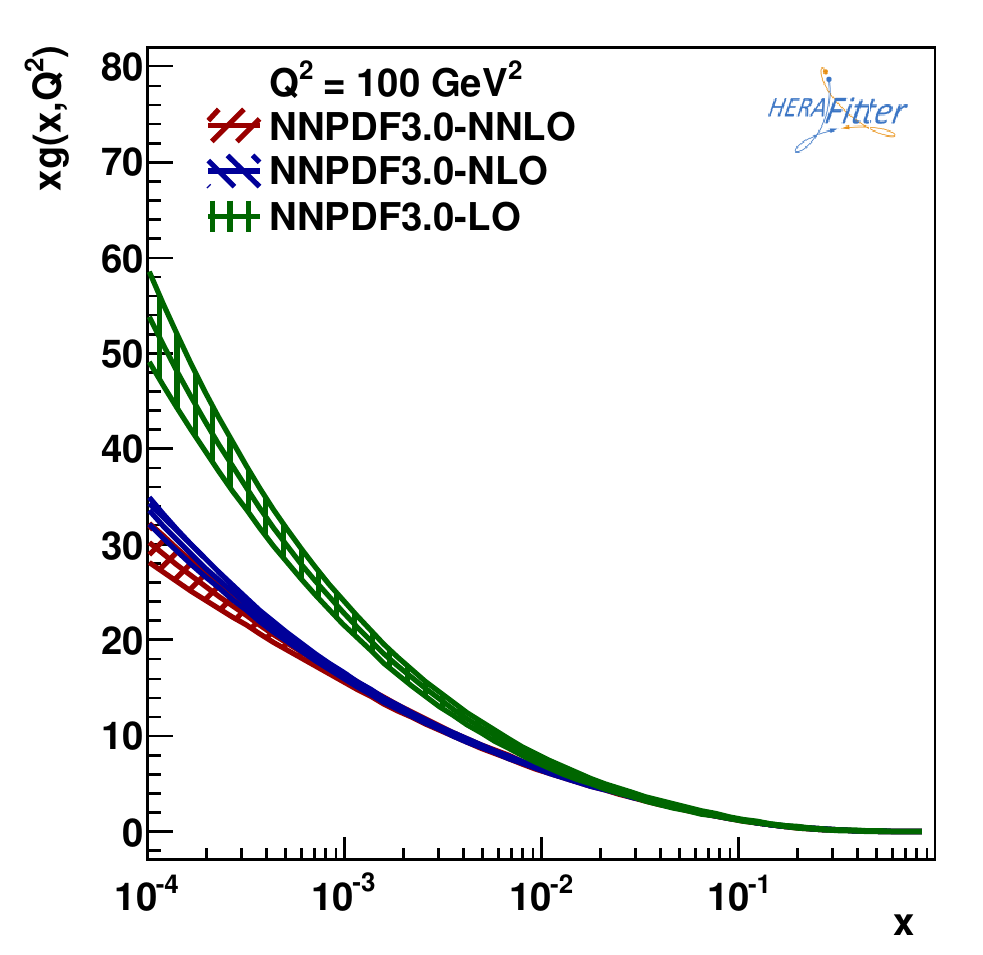}
\caption{\it   Proton's structure 
 as a function of  
 momentum fraction 
$x$: gluon density 
 at different  mass  scales  $Q^2$~\protect\cite{Ball:2014uwa}.}
\label{fig:REFfigstrfun}
\end{center}
\end{figure}

A second example 
concerns  the rise of  
proton's   structure functions 
at small 
longitudinal momentum fractions. 
Since in 
$pp$ collisions the product of 
initial-state  
longitudinal   fractions  scales 
like $1 / s $ at fixed momentum 
transfer, where 
$s$ is the squared 
centre-of-mass energy, 
as we push forward the 
high-energy frontier  
more and more 
events at small 
longitudinal fractions  contribute 
to processes 
probing  short-distance physics. 
Many hard-production cross sections 
at the LHC  receive  sizeable contributions 
from proton's structure functions in 
this region. As parton longitudinal momenta 
become small, the fraction of momentum 
carried by transverse degrees of  freedom 
 becomes increasingly important.

Fig.~\ref{fig:REFfigstrfun}       
shows 
the proton's gluon density 
resulting from global fits~\cite{Ball:2014uwa}   
to hadronic collision data, 
performed at LO, 
NLO,   NNLO~\cite{Vogt:2004mw,Vermaseren:2005qc,Moch:2004xu} 
of perturbation 
theory, 
   as a function of the 
longitudinal 
momentum fraction $x$ for 
different values of the evolution 
mass scale $Q^2$.  
In the low-$x$  regime 
 the  perturbative 
higher-order corrections  
to structure functions 
 are  large, and 
  the gluon pdf uncertainty is 
large.  The 
 strong 
 corrections 
at low $x$  
come from multiple  
radiation of gluons 
over long  
intervals in rapidity~\cite{Gribov:1984tu,Mueller:1993rr}, 
in regions not ordered in 
the gluon transverse 
momenta $p_T$, 
and are present 
beyond NNLO to all orders of 
perturbation theory~\cite{Catani:1993rn,Catani:1994sq}.  The 
theoretical framework to resum 
these unordered 
multi-gluon emissions 
is   a generalized form   of 
QCD factorization~\cite{Catani:1990xk,Catani:1990eg} 
in terms of TMD pdfs. 
Analogously to the Drell-Yan 
case discussed earlier,  the 
TMD pdfs obey a suitable set of 
 evolution equations~\cite{Kuraev:1976ge,Kuraev:1977fs,Balitsky:1978ic},   appropriate 
to this kinematic region.  These  
provide another generalization, 
valid in the high-energy limit,  
of the ordinary 
renormalization-group evolution.  
  The TMD factorization in this case 
allows one to resum logarithmically 
enhanced corrections in the 
ratio $\sqrt{s}  / Q  $  
to  all higher orders in the   QCD coupling.

Besides the above  examples 
of Drell-Yan and structure 
functions, 
 TMD factorization theorems 
apply to a wide variety of processes 
at the LHC. In particular,  with 
extensive measurements of  
Higgs boson production at the LHC Run II,  a 
new set of QCD processes becomes  
available in which the Higgs 
boson acts as  a color-singlet, 
pointlike source  (in the 
heavy top limit) which   couples to 
gluons. This is to  be 
contrasted  with   
Drell-Yan and deep-inelastic 
scattering  cases,  
 based on 
weak and electromagnetic 
currents  providing  color-singlet 
pointlike sources coupled  to 
quarks. This  opens up the 
possibility of a new program of 
precision QCD measurements 
in gluon fusion at high mass scales 
in the LHC high-luminosity 
runs~\cite{Cipriano:2013ooa,Haevermaet:2014sda}.

Analogously to the case of 
vector bosons in the example  
of Fig.~\ref{fig:REFfigZ},  
 theoretical predictions for 
the  Higgs-boson production 
 differential spectrum 
over the whole range in transverse 
momenta accessible at the LHC 
require 
generalized QCD factorization,    
based on  initial-state gluon distributions 
that include polarization and 
transverse-momentum  
degrees of freedom. Compared to the vector boson  case, however, 
new features arise which  are 
 associated with the role of 
gluon polarizations in gluon-gluon  
scattering.

\begin{figure}[htbp]
\begin{center}
\includegraphics[scale=0.67]{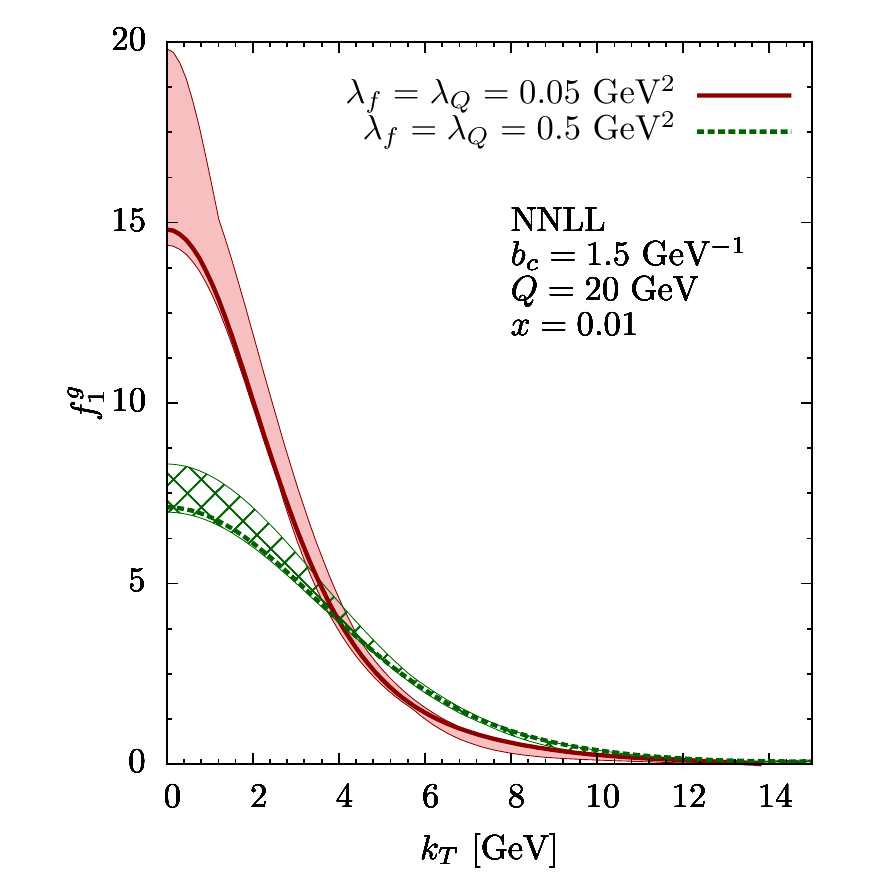}
\includegraphics[scale=0.67]{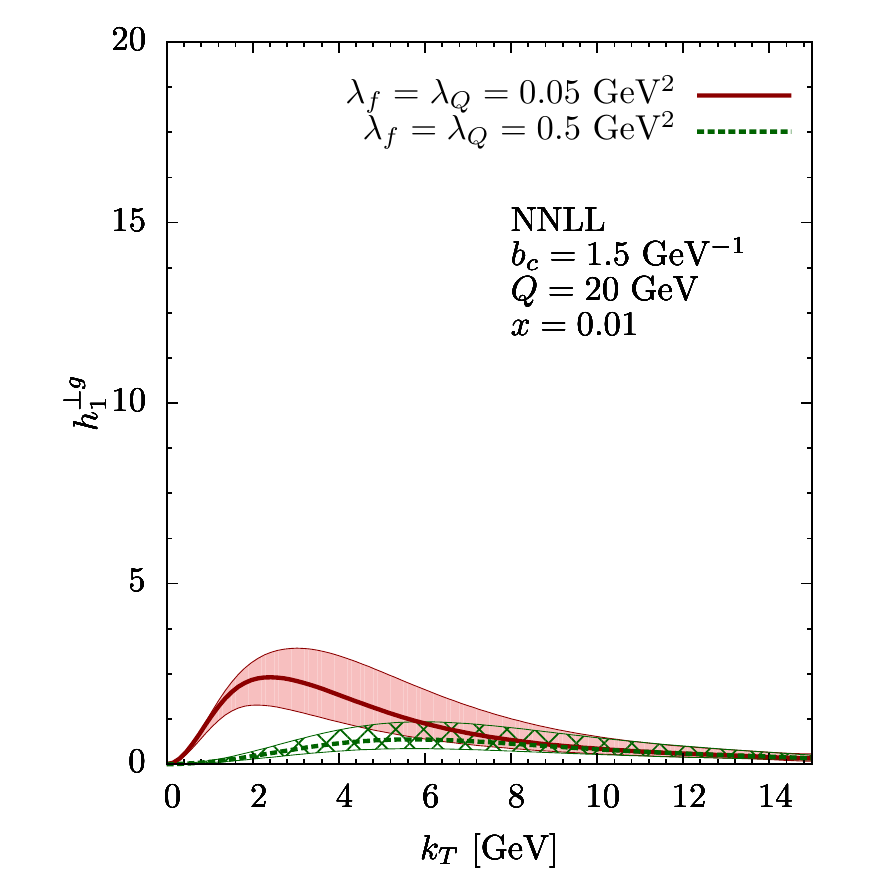}
\caption{\it  The transverse momentum 
dependence of the unpolarized  (left) and  
 linearly polarized (right) 
gluon  distributions~\protect\cite{Echevarria:2015uaa} contributing to 
the gluon-fusion Higgs production  spectrum. The results are plotted 
for evolution scale $ Q = 20 $ GeV 
and longitudinal momentum fraction 
$ x = 0.01 $, and  for different values of the 
nonperturbative parameters discussed in~\protect\cite{Echevarria:2015uaa}.The 
red and green bands around each curve 
correspond to variations by  factor  2 
of the resummation scale and rapidity scale in the 
calculation~\protect\cite{Echevarria:2015uaa}.} 
\label{fig:REFh-to-f}
\end{center}
\end{figure}

More precisely, in the high-energy 
limit $ \sqrt{s} \gg m_H$ the Higgs boson production 
from gluon fusion 
is dominated by a single eikonal  gluon 
polarization~\cite{Hautmann:2002tu}.  
The contribution of this 
polarization depends on 
the gluon transverse momentum 
and can be rewritten 
in terms of the high-energy 
projection operator  defined  
in~\cite{Catani:1994sq}. 
A complete set of operators for 
polarization dependent 
and transverse momentum dependent 
gluon distributions is given 
in~\cite{Mulders:2000sh}.  
In the region of low 
Higgs-boson transverse 
momenta, $ q_T \ll m_H $,  
the contributions of polarized 
gluons to the Higgs spectrum 
have been studied 
both 
perturbatively~\cite{Mantry:2009qz,Mantry:2010mk,Catani:2010pd,Becher:2012yn,Nadolsky:2007ba}    
 and nonperturbatively~\cite{Boer:2011kf,Boer:2013fca,Sun:2011iw,Boer:2014tka,Echevarria:2015uaa}. 
An  example is shown in  Fig.~\ref{fig:REFh-to-f}~\cite{Echevarria:2015uaa}, 
where the  unpolarized and 
 linearly polarized    gluon distributions contributing to the 
Higgs boson spectrum at small $q_T$ 
are  plotted as a function of transverse 
momentum.  
 The presence  of polarized gluon components 
(even in unpolarized beams) characterizes 
 gluon fusion processes 
and has no analogue in the Drell-Yan case. 
In particular the component  in the 
right hand side plot 
of Fig.~\ref{fig:REFh-to-f} 
is  a gluon TMD 
distribution with double spin flip (see Table~II 
ahead, top right corner).  
From the point of view of 
perturbative power counting, double 
 spin flip effects start to contribute to the 
Higgs $q_T$ spectrum at the NNLO (but may  
contribute earlier in more complex,  
less inclusive observables associated with 
Higgs production). Detailed   
measurements of  Higgs boson  
final states  will  allow the  
QCD dynamics of polarized gluons and 
their  correlations  to be 
explored  experimentally for the first time.

\begin{figure}[htbp]
\begin{center}
\includegraphics[scale=0.55]{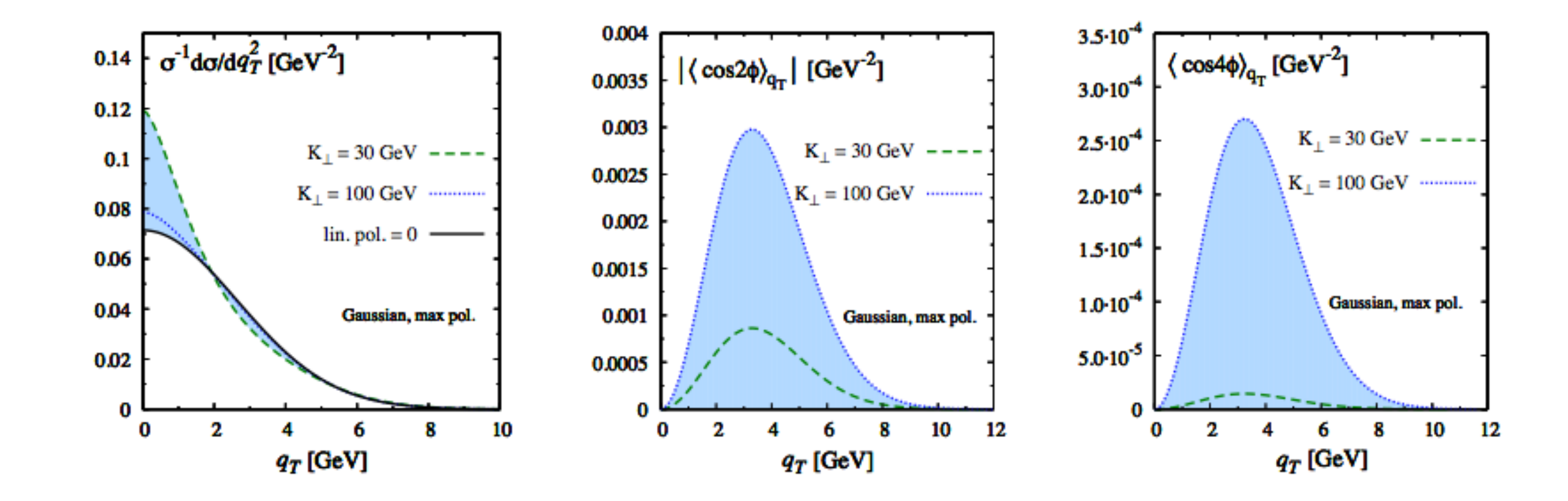}
\caption{\it  Theoretical predictions~\protect\cite{Boer:2014lka}
for the 
transverse momentum 
distribution (left), $\cos 2 \phi$ asymmetry (middle) and $ \cos 4 \phi$ asymmetry (right) in  Higgs boson + jet 
production at small   transverse 
momenta $q_T$  of the  Higgs + jet pair. Here 
$K_\perp$ represents the average 
of the Higgs and jet transverse momenta, and 
 the shaded blue areas represent the range of 
the asymmetries as $K_\perp$ varies from 0 to 
$\infty$.} 
\label{fig:REFhiggsasymm}
\end{center}
\end{figure}

For both the Drell-Yan and Higgs cases, 
in addition to the inclusive spectra 
an extensive experimental program 
at the LHC is  devoted to  
 the associated production of heavy bosons with jets.  
The region in which the boson and 
leading jet are nearly back-to-back
presents  features comparable 
to the 
discussion given above for the 
low-$q_T$ part of the inclusive 
spectra. For instance, 
a study of TMD 
gluon  contributions to 
Higgs + jet final states    
in  which 
the imbalance between the 
boson and 
leading-jet transverse momenta  is 
small    is   reported  in 
 Fig.~\ref{fig:REFhiggsasymm}~\cite{Boer:2014lka},   
showing  the boson-jet pair's 
 transverse momentum distribution 
and  azimuthal asymmetries.

The case of  associated boson + jet production    when  
 the imbalance 
between the 
boson and 
leading-jet transverse momenta
 is not small, on 
the other hand,   probes   
 the physics of final states with 
multiple  jets.  The role of TMD 
parton distributions in scenarios  
with  high 
jet multiplicity is discussed in the 
next section, and  
serves to   illustrate the 
connection  of   TMDs with  the 
 kinematic 
region of large transverse momenta.

\begin{figure}[htbp]
\begin{center}
\includegraphics[scale=0.55]{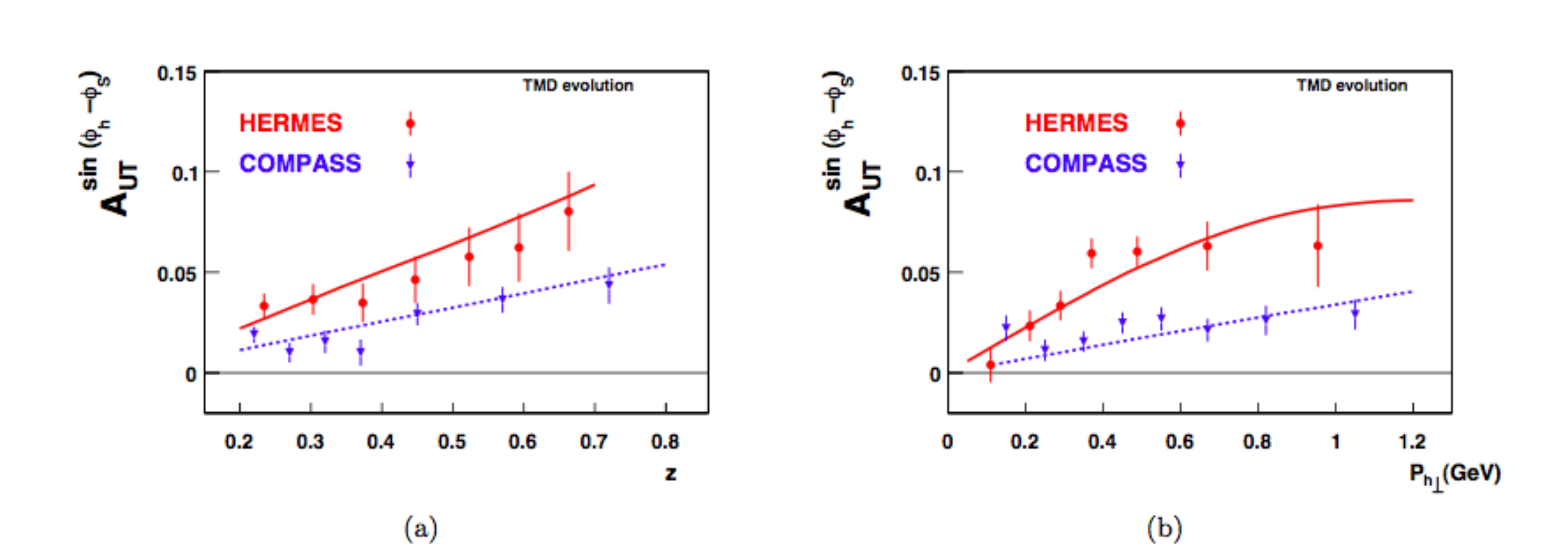}
\caption{\it  Sivers asymmetry 
measurements~\protect\cite{Airapetian:2009ae,Bradamante:2011xu}   and fits~\protect\cite{Aybat:2011ta}
as a function of hadron's longitudinal  
momentum  fraction (left) and  
transverse momentum (right).} 
\label{fig:sivers}
\end{center}
\end{figure}

An extension of the 
 methods discussed above  
for Drell-Yan and Higgs production 
applies 
 to the transverse momentum 
spectra of heavy flavor pairs, e.g. top quarks. 
Unlike the 
case of color-singlet  currents 
coupled to quarks (as in 
Drell-Yan production) 
or gluons (as in Higgs boson 
production in the heavy top limit), 
 heavy-quark pair production 
constitutes a composite non-pointlike 
probe,  containing    
 color-charged particles  in the 
lowest-order final state and receiving  
contribution from 
 both quark and gluon TMD channels.   
Color correlations over long timescales between initial    
and final states will break factorization in the region of very 
small transverse-momentum  imbalance of the   pair~\cite{Zhu:2012ts,Li:2013mia,Zhu:2013yxa,Catani:2014qha,Rogers:2010dm,Rogers:2013zha,Collins:2007nk,Vogelsang:2007jk}. Studies of this region and the interplay of perturbative and nonperturbative 
contributions will help understand quantitatively these effects.

Another  area  for 
 applications of TMDs 
  concerns 
single spin asymmetries and 
azimuthal asymmetries in 
polarized collisions. A classic 
example is the Sivers 
asymmetry~\cite{Sivers:1989cc,Brodsky:2002cx,Collins:2002kn}. 
Fig.~\ref{fig:sivers}~\cite{Aybat:2011ta}
shows low-energy 
measurements~\cite{Airapetian:2009ae,Bradamante:2011xu} of the 
Sivers transverse single spin asymmetry 
along with  results of the fit~\cite{Aybat:2011ta}.  
For hadron's  transverse 
momenta  sufficiently 
 small compared to 
the virtuality scale $Q$ of the 
deep inelastic (or Drell-Yan) process,  
  spin asymmetries   obey TMD factorization formulas of the same 
kind~\cite{Collins:2011zzd}  discussed 
above for the unpolarized case of 
low $q_T$ Drell-Yan.  A  combined  understanding of 
current high-energy unpolarized measurements and low-energy 
spin asymmetry measurements  is important for the planning of 
 future polarized collider~\cite{Aschenauer:2015eha,Accardi:2012qut} and 
fixed-target~\cite{Lansberg:2014myg,Brodsky:2012vg} experiments.

\begin{table}
\includegraphics[scale=0.3]{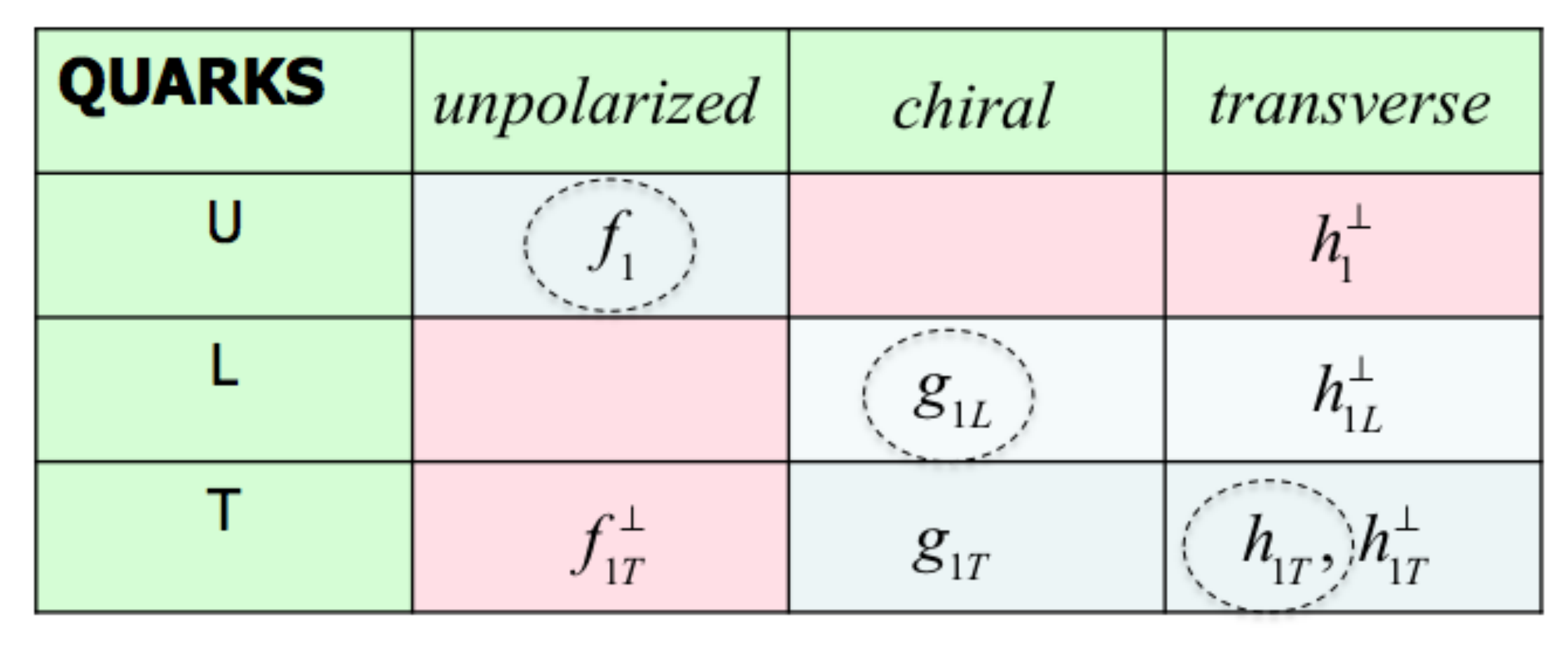}
\caption{\it  
Quark TMD pdfs: columns represent quark polarization, rows represent  hadron polarization. 
Distributions encircled by a dashed line are the  ones
which survive integration over transverse momentum. The shades of the boxes
(blue versus pink) indicate structures that are T-even or T-odd,
respectively.  T-even and T-odd structures involve, respectively,
an even or odd number of spin flips.}
\label{t:quark_tw2_tmdpdfs}
\end{table}

We conclude this section by presenting the 
full leading-twist  set of   
polarization dependent 
and transverse momentum 
dependent 
 parton densities in a spin-1/2 hadron. 
These are shown in Table~I and 
Table~II, for the 
quark~\cite{Mulders:1995dh,Bacchetta:2006tn} 
and  
gluon~\cite{Mulders:2000sh,Meissner:2007rx}    
cases respectively, 
including the distributions in  
 unpolarized hadrons (top rows), 
longitudinally polarized 
hadrons (middle rows), 
transversely polarized 
hadrons (bottom rows). 
(See~\cite{Ralston:1979ys,Barone:2001sp,Idilbi:2004vb,Anselmino:1994tv,Anselmino:1995vq,Anselmino:1996qx,Anselmino:2005sh}
for slightly different classifications.) 
Gauge-invariant operator 
 definitions may be given 
for each of 
the TMD distributions in terms of  
 nonlocal operator combinations,  
in which appropriate Wilson-line gauge links 
are associated with 
quark and gluon 
fields~\cite{Collins:2011zzd,Bomhof:2007xt,Dominguez:2011wm,Buffing:2012sz,Buffing:2013kca,Boer:2015kxa}. 
Operator definitions are  instrumental in analyzing both factorization and 
potential sources  of factorization breakdown, and in setting  up lattice 
  calculations~\cite{Musch:2010ka,Musch:2011er,Ji:2014hxa,Ma:2014jga}  
 of parton distributions.

\begin{table}
\centering
\includegraphics[scale=0.3]{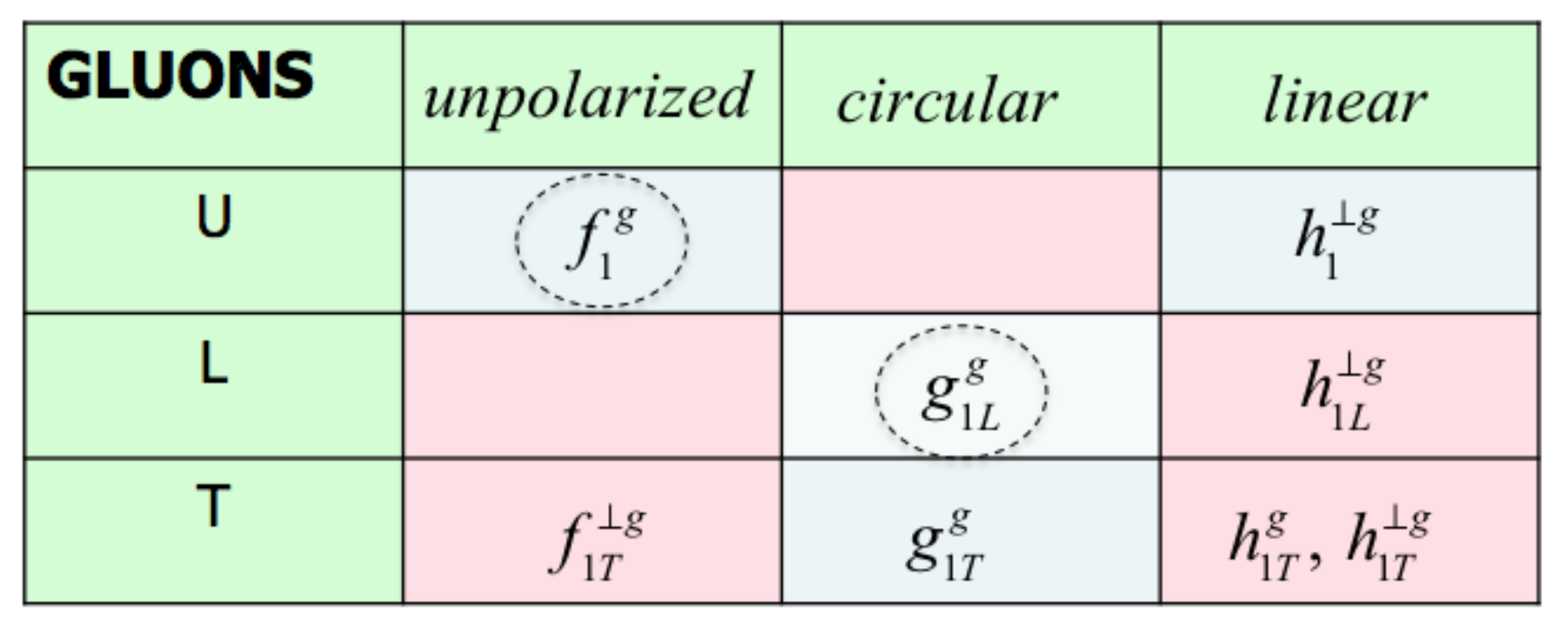}
\caption{\it  
Gluon TMD pdfs: columns represent gluon polarization, rows represent  hadron polarization. 
Distributions encircled by a dashed line are the  ones
which survive integration over transverse momentum. The shades of the boxes
(blue versus pink) indicate structures that are T-even or T-odd,
respectively.  T-even and T-odd structures involve, respectively,
an even or odd number of spin flips.   Linearly polarized gluons
represent a double spin flip 
structure.}
\label{t:gluon_tw2_tmdpdfs}
\end{table}

\section{TMDs and large  transverse momenta}

Unlike the 
low-$q_T$ Drell-Yan factorization 
theorem~\cite{Collins:1982wa,Collins:1984kg,Collins:2011zzd} 
 and its extensions for gluon fusion processes, 
the high-energy factorization 
theorem~\cite{Catani:1994sq,Catani:1990eg,Catani:1990xk,Catani:1993ww} 
is valid for arbitrarily large 
momentum transfer. 
It  is 
 based on the high-energy expansion 
$\sqrt{s} \to \infty$ and can be 
applied in the ultraviolet region of 
high  $q_T$.    
  It allows one, for  
example,  to obtain the structure of logarithmic scaling violations in 
DIS at high energy   (see~\cite{Vogt:2004mw,Vermaseren:2005qc,Moch:2004xu})   and 
to resum    logarithmic corrections  
 of higher order in $\alpha_s$  to 
Higgs and top quark production 
cross sections   (see~\cite{Luisoni:2015xha,Czakon:2015xqa,Czakon:2013goa}).   
In this section we apply this 
 theorem to discuss  the role of 
TMDs in the region of 
perturbative transverse momenta, 
in particular in 
the high-$q_T$ part 
of the Drell-Yan spectrum in 
Fig.~\ref{fig:REFfigZ}. 

The basic observation is that 
 the LHC kinematics leads 
to copious production of 
final states 
 in which 
a high-$q_T$ vector boson recoils  
against  multiple hard jets.  Ref.~\cite{Dooling:2014kia} studies 
$W$-boson + $n$ jets 
final states using 
 TMD high-energy  factorization~\cite{Catani:1990eg}. 
The motivation for this is 
twofold: a) kinematical: it has  recently been pointed  out~\cite{Dooling:2012uw,Dooling:2013rta,Hautmann:2012dw} 
 that  collinearity approximations,
 once combined with energy-momentum conservation constraints,  
give rise to longitudinal momentum 
shifts  and  sizeable 
showering corrections in the Monte 
Carlo  algorithms used to simulate multi-jet final states at the LHC; 
b) dynamical:  it has long been 
known~\cite{Ciafaloni:1987ur,Marchesini:1992jw,Hautmann:2008vd} 
that, when 
the picture of multi-jets from 
finite-order perturbative matrix 
elements matched with 
collinear parton showers 
is pushed to higher and higher 
energies, new effects arise in jet 
multiplicity distributions and 
angular correlations due to soft but 
finite-angle multigluon radiation. 
Both these kinematical and dynamical 
effects can be taken into account by 
 a TMD treatment of 
 QCD parton shower  evolution~\cite{Hautmann:2008vd}.

\begin{figure*}
\centering
\includegraphics[scale=0.4]{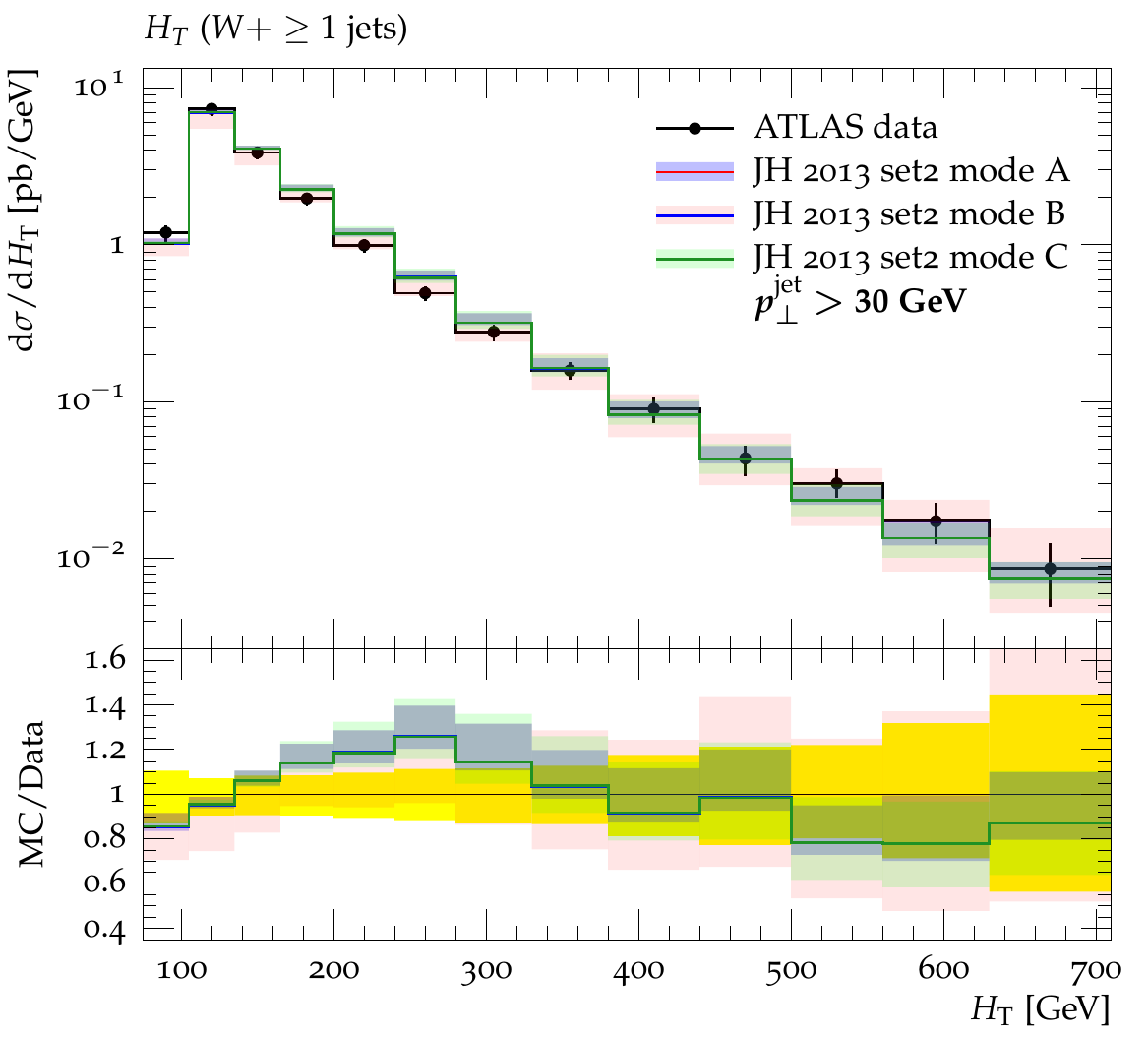}
\includegraphics[scale=0.4]{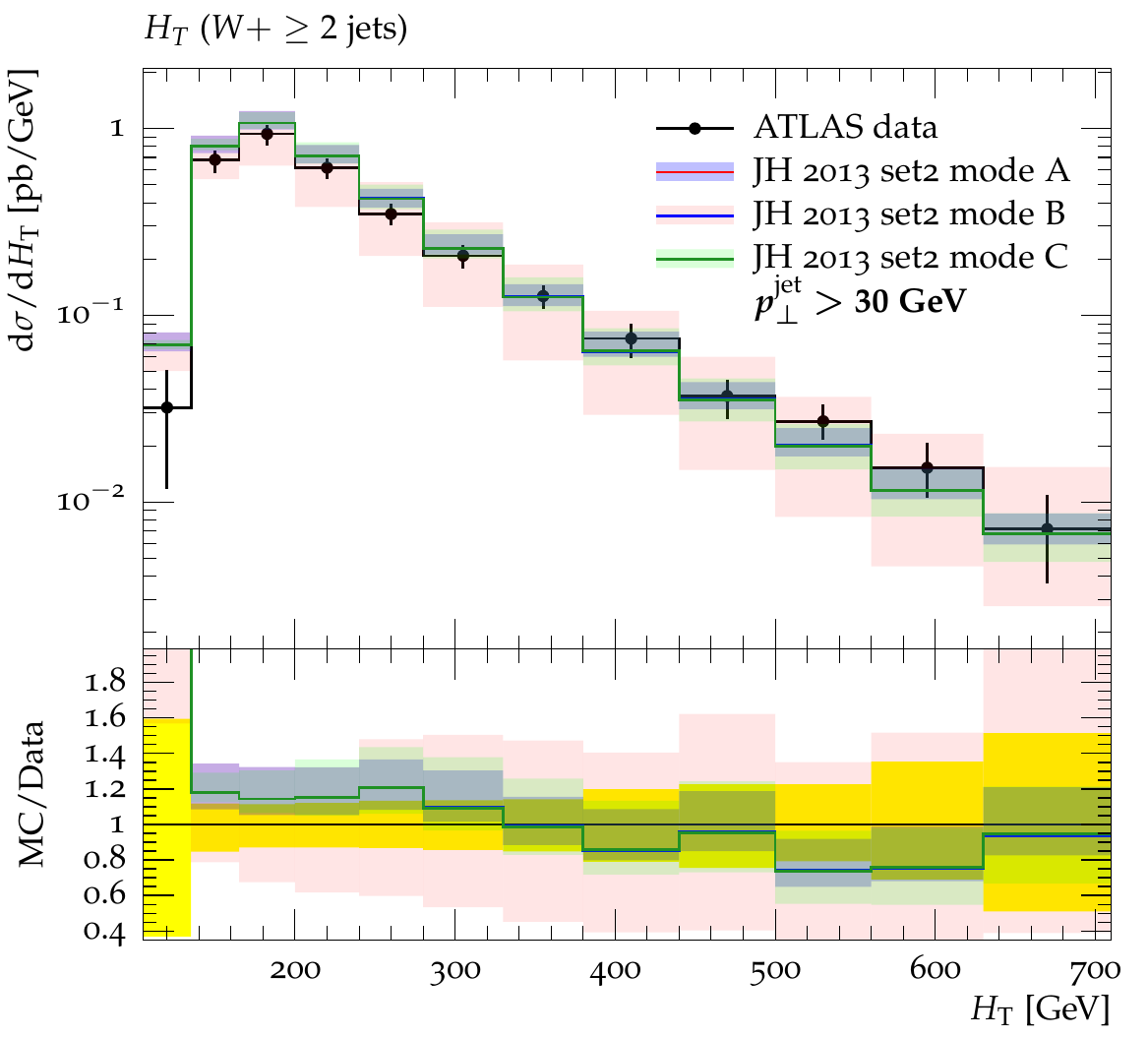}
\includegraphics[scale=0.4]{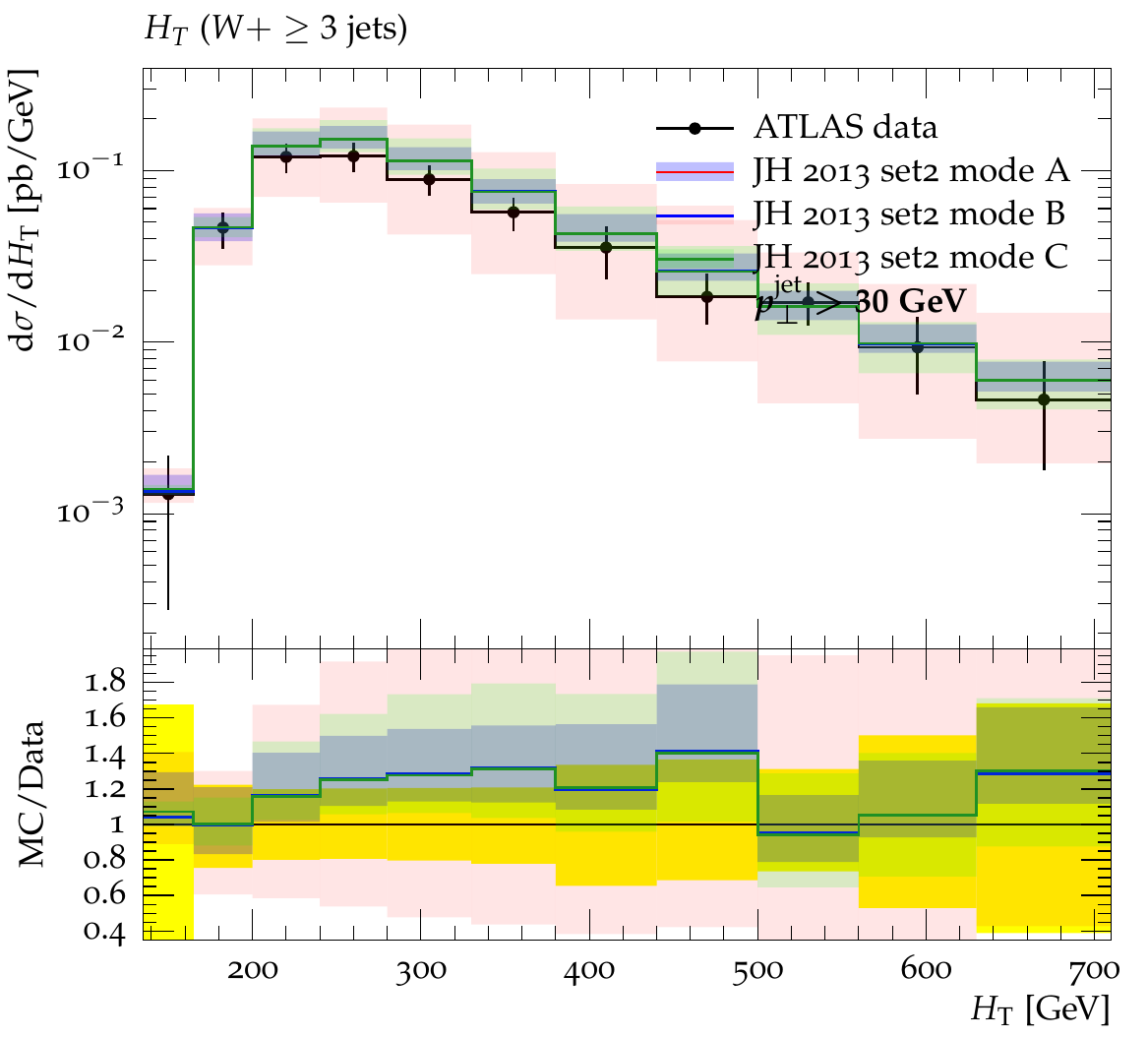}
  \caption{\it Total transverse energy $H_T$ distribution 
  in  final states   
with 
$W $-boson + $n$  jets  at the LHC, for 
 $ n \geq 1 $ (left),  $ n \geq 2 $ (center), $ n \geq 3 $ (right). 
The purple, pink and green bands 
correspond to the different methods 
 described in~\protect\cite{Dooling:2014kia} to estimate 
 theoretical uncertainties.  The experimental data are 
 from~\protect\cite{Aad:2012en}, with the  
experimental uncertainty represented by  the yellow band.}
\label{fig:Ht}
\end{figure*}

To achieve this, 
 Ref.~\cite{Dooling:2014kia} uses the 
exclusive formalism of CCFM 
evolution 
equations~\cite{Ciafaloni:1987ur,Catani:1989sg,Marchesini:1994wr} 
implemented in~\cite{Hautmann:2014uua}. 
The TMD pdfs to which  evolution is applied are determined 
from fits to 
the precision DIS 
data~\cite{Hautmann:2013tba}. 
 By evolving these TMD pdfs up to 
the scale of $W$ +  jets and 
coupling them with appropriate,  
perturbatively calculated 
high-energy  matrix 
elements,  one obtains predictions 
for $W$-boson + $n$ jets observables.  
Fig.~\ref{fig:Ht} shows the total 
transverse energy $H_T$ distribution 
in final 
states with 
$ W $-boson + $n$  jets, with $ n = 1 , 2 , 3 $, at the 
LHC.  For comparison  the experimental 
measurements~\cite{Aad:2012en} 
(jet rapidity $|\eta |  < 4.4$, jet transverse momentum 
$p_T >   30 $ GeV) are plotted. 
The uncertainty bands on the 
 theoretical predictions are 
 described  in~\cite{Dooling:2014kia}, and 
largely reflect  uncertainties on 
TMDs determinations, estimated  according to  three different 
approaches corresponding to the 
three color bands. 

\begin{figure*}
\centering
\includegraphics[scale=0.4]{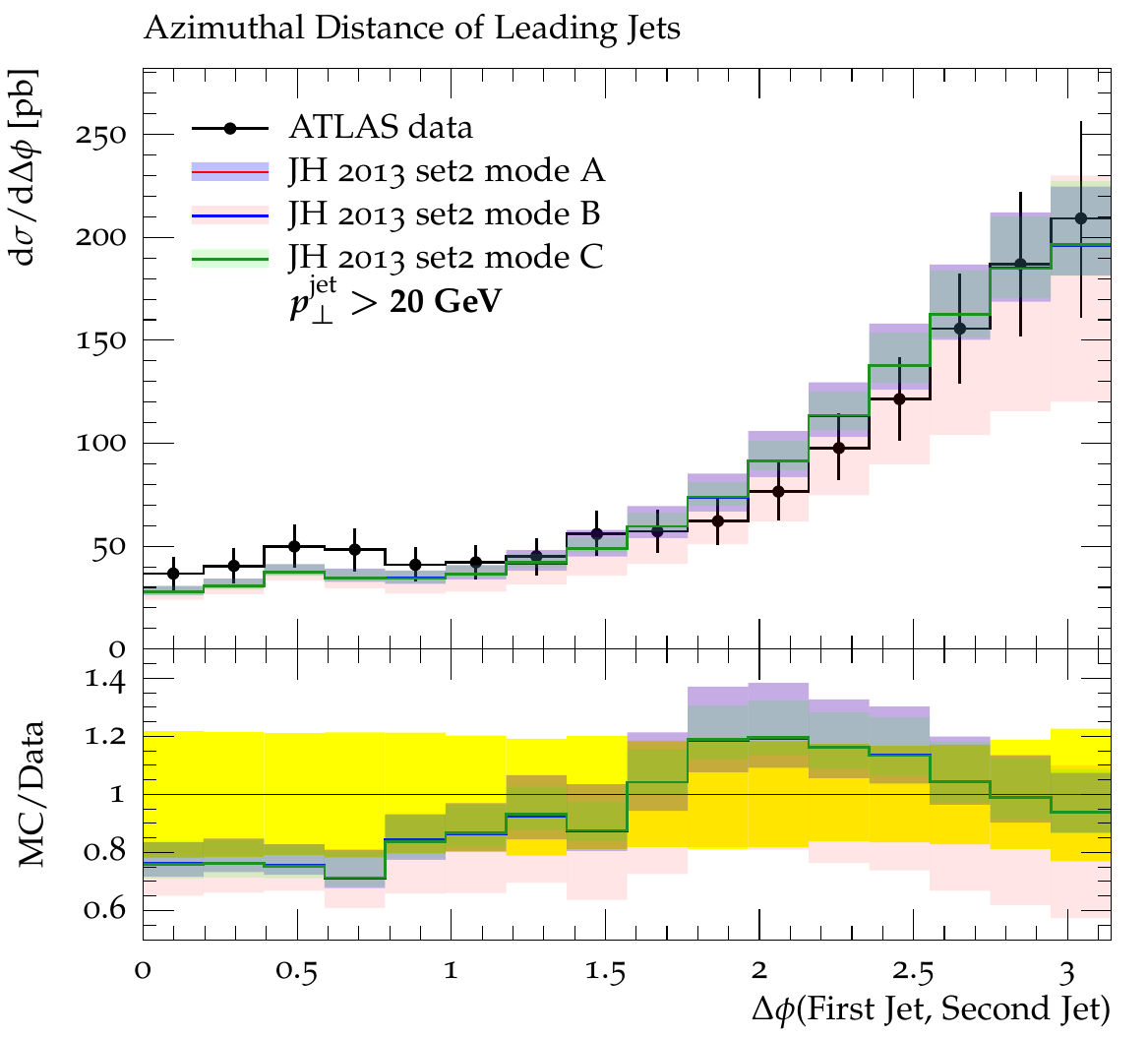}
\includegraphics[scale=0.4]{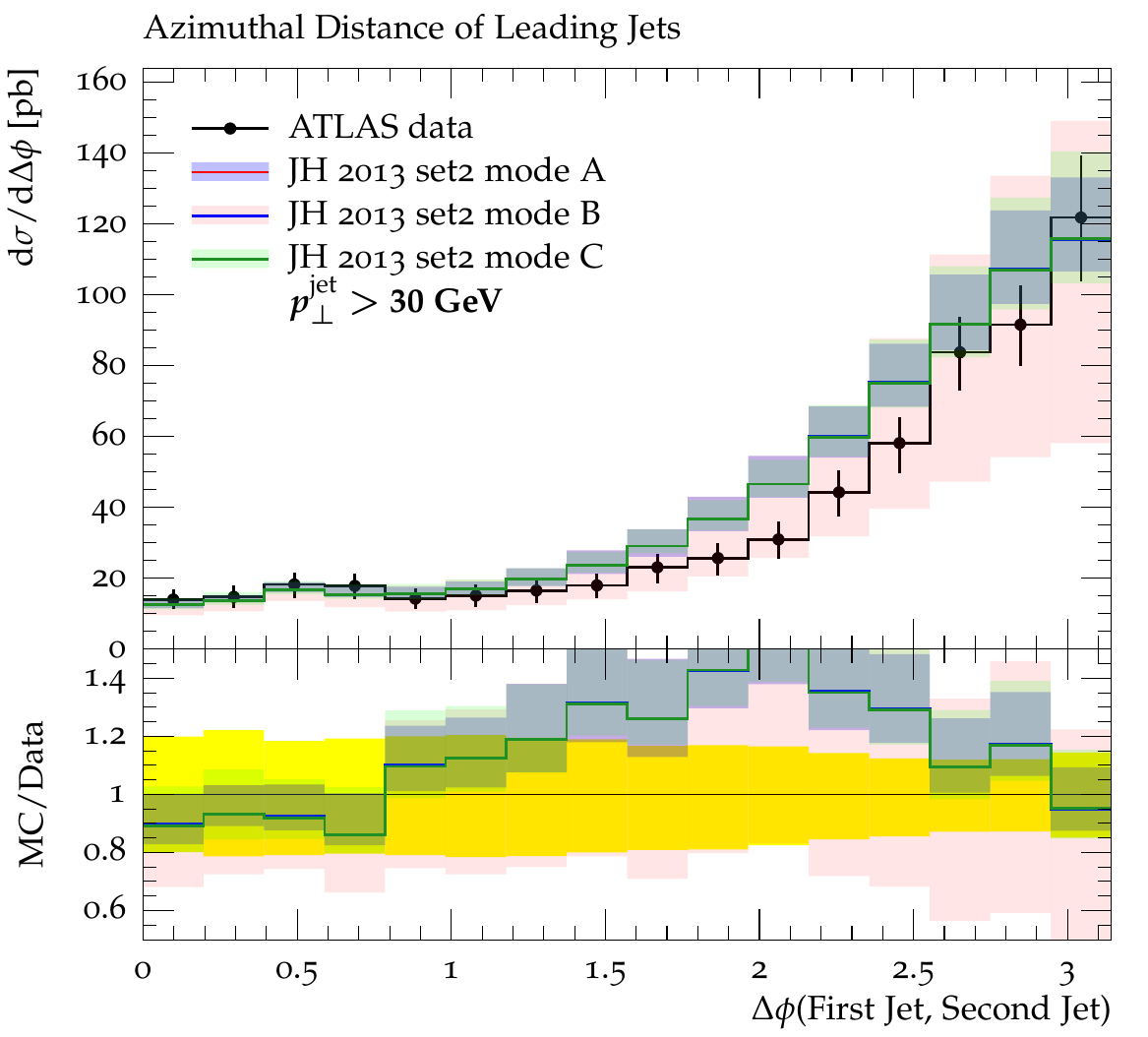}
\includegraphics[scale=0.4]{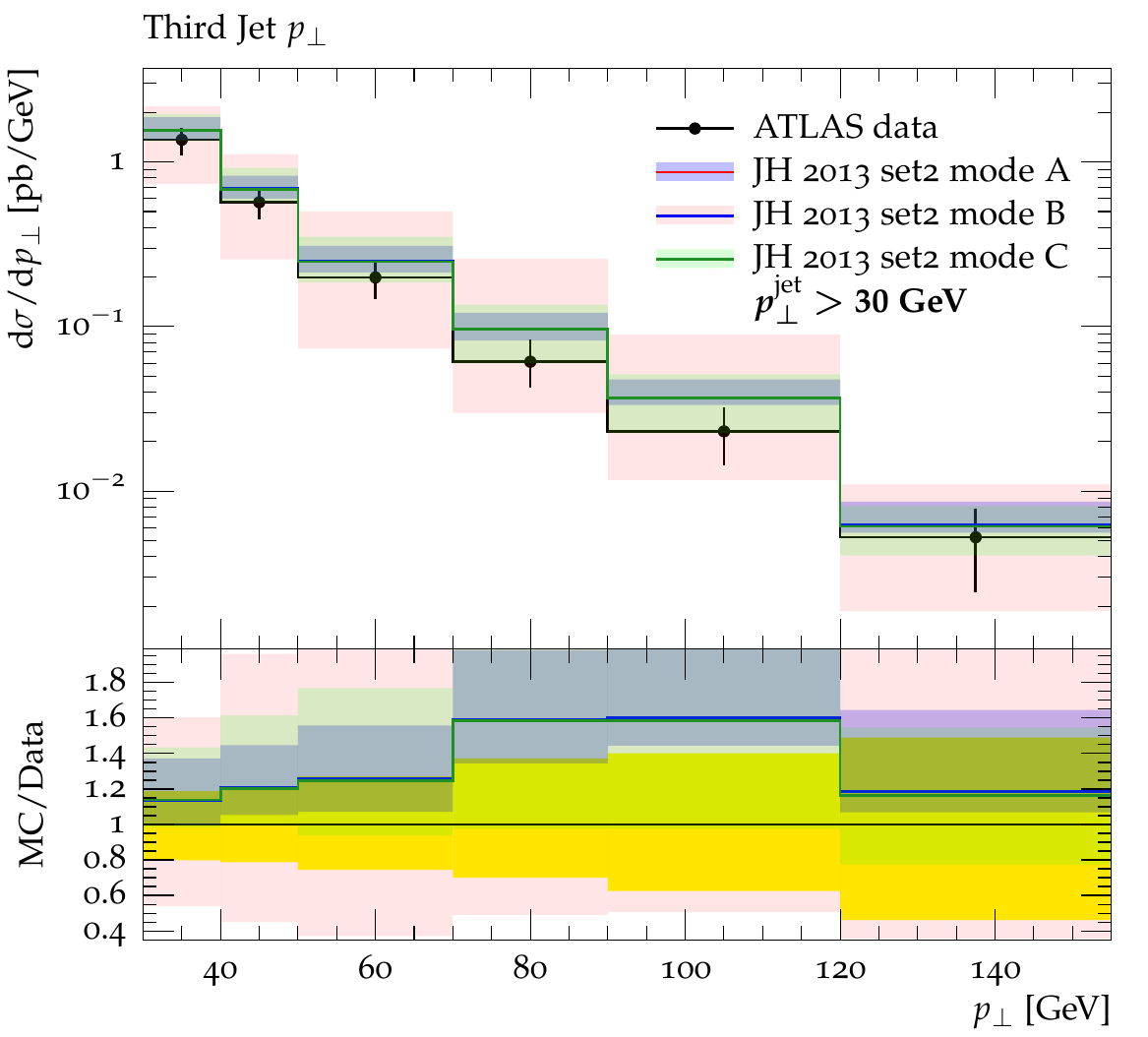}
  \caption{\it  Azimuthal 
correlation of the two 
leading jets 
associated    with 
$W $-bosons, for $ p_T > 20 $ GeV (left) and 
  $ p_T > 30 $ GeV (center), and   
 transverse momentum  of the 
third jet ($ p_T > 30 $ GeV) (right).
The purple, pink and green bands 
correspond to the different methods 
 described in~\protect\cite{Dooling:2014kia} to estimate 
 theoretical uncertainties.  The experimental data are 
 from~\protect\cite{Aad:2012en}, with the  
experimental uncertainty represented by  the yellow band.
}  
\label{fig:azim}
\end{figure*}

The TMD high-energy factorization  
predicts  azimuthal correlations 
in the $W$ + multi-jet final states. 
Fig.~\ref{fig:azim} shows results 
for the azimuthal correlation between the two leading jets, along with the   
transverse momentum of the third jet.

Current limitations of 
the approach described  
above  and ongoing improvements 
  are  discussed 
in~\cite{Dooling:2014kia,Hautmann:2015koa,Hautmann:2014dya}, 
and 
 include in particular the 
treatment of TMD quark density 
distributions, and the accuracy of 
determinations of the  gluon density 
distribution over the whole range of  
 longitudinal  momentum fractions $x$ relevant to the LHC kinematics. 
 The results in 
Figs.~\ref{fig:Ht} 
 and \ref{fig:azim}    
 are  however encouraging,  and sufficiently general, 
in the context of   approaches  that  aim  
to go beyond fixed-order perturbation theory and 
appropriately    take  account  of   nonperturbative effects. 

As TMDs describe    nonperturbative transverse momentum 
 dynamics  in the hadron, 
  they    
may  provide a  suitable  framework  not only for the factorization of the 
hard process  but also  to 
incorporate effects from  
  soft particle production and 
multi-parton 
interactions~\cite{Diehl:2014vaa,Hautmann:2012xa}.

\begin{figure}[htbp]
\begin{center}
\includegraphics[scale=0.3]{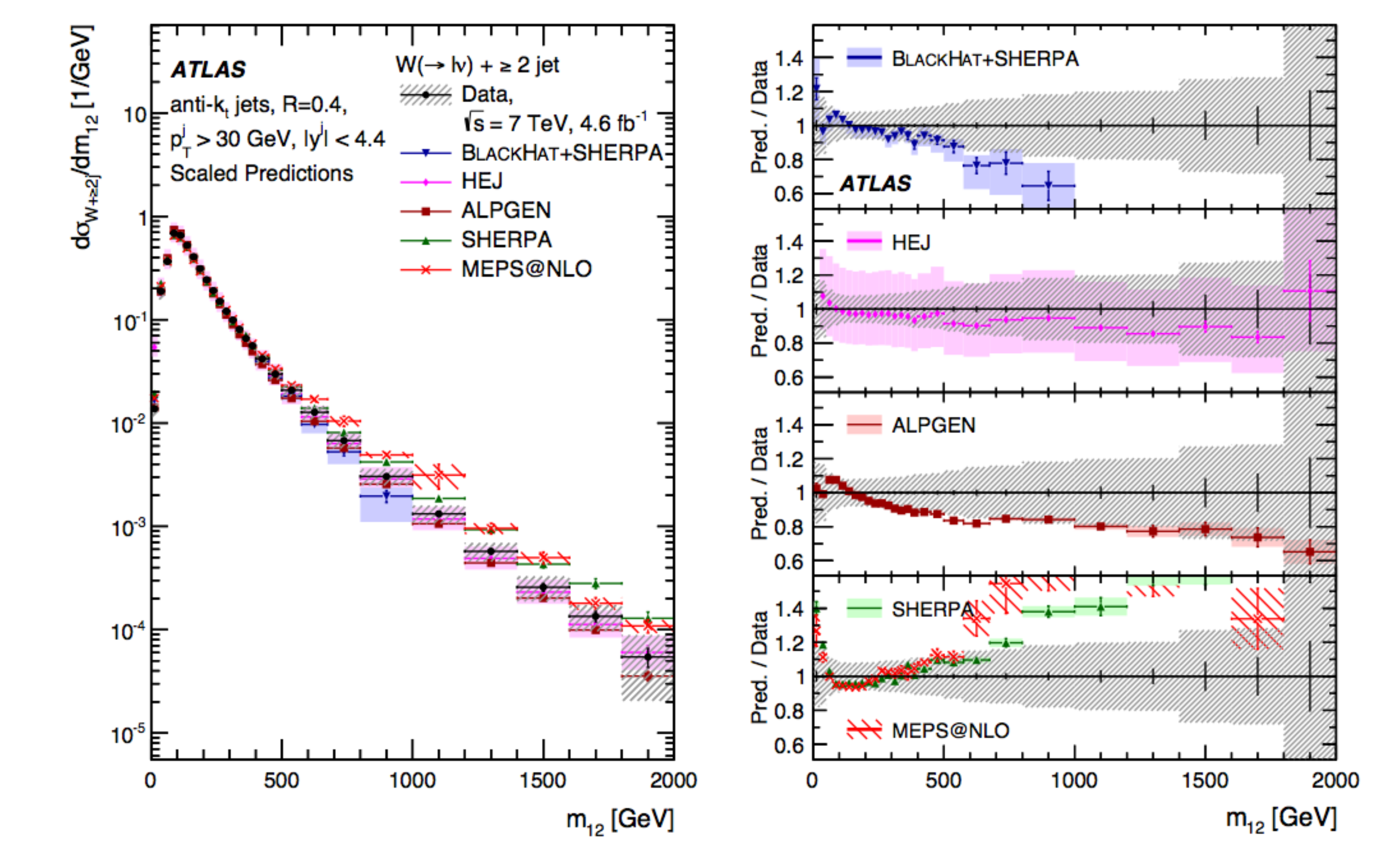}
\caption{\it Di-jet  invariant mass  measured~\protect\cite{Aad:2014qxa} 
in LHC  final states   
with $W $-boson + $2$  jets, compared 
with  parton-shower Monte Carlo calculations.}
\label{fig:REFatlasm12}
\end{center}
\end{figure}

\begin{figure}[htbp]
\begin{center}
\includegraphics[scale=0.6]{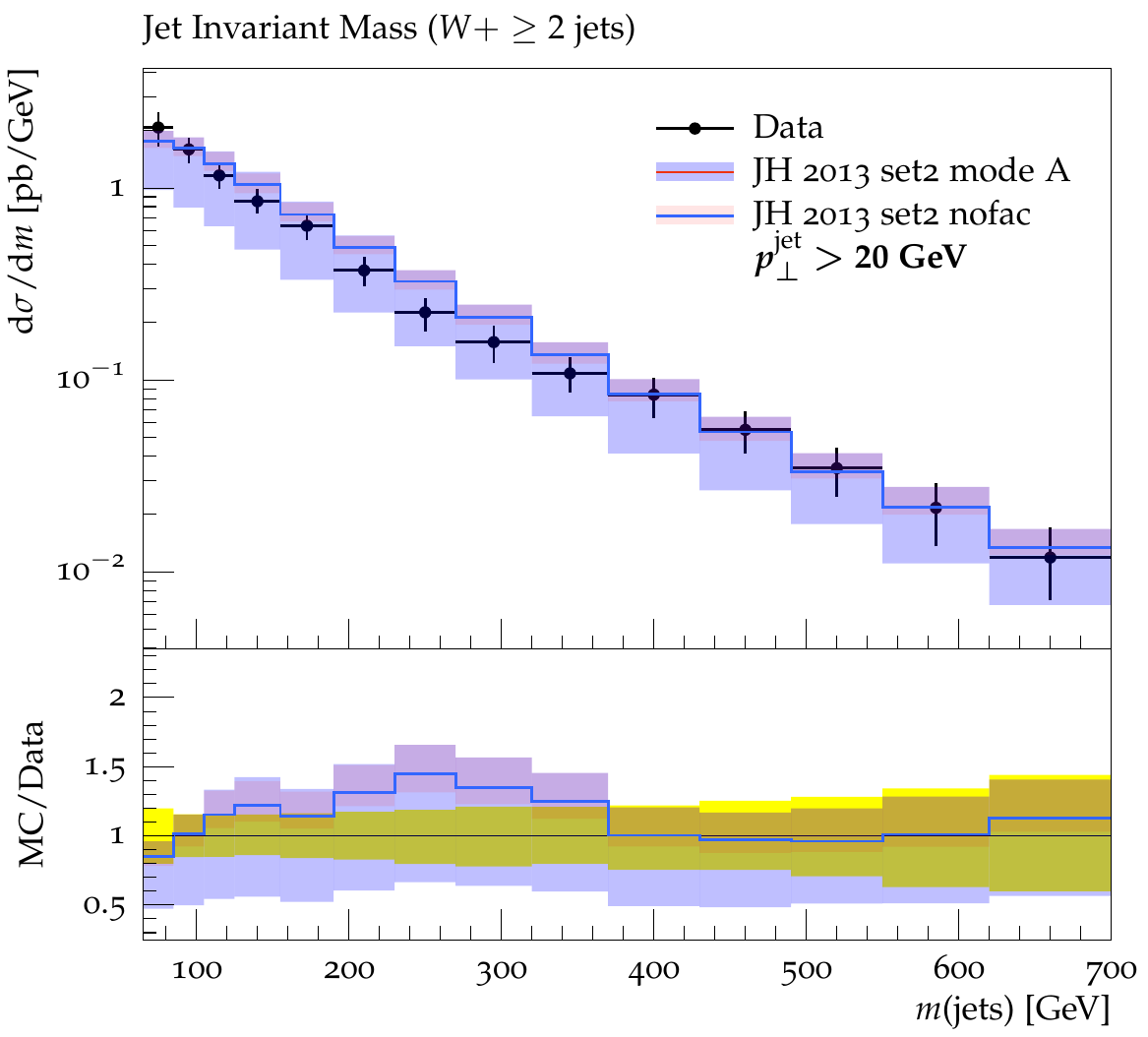}
\caption{\it Di-jet  invariant mass  
distribution in    
 $W $-boson + $2$  jets final states, 
computed from the 
approach~\protect\cite{Dooling:2014kia}. 
The experimental data are 
from~\protect\cite{Khachatryan:2014uva}.}
\label{fig:REFm12}
\end{center}
\end{figure}

It is worth noting  that 
while for sufficiently inclusive observables in 
$W$ + jets production 
calculations based on 
collinear parton showers matched with 
finite-order perturbative matrix elements 
describe   measurements 
at Run I very well, this   may   not 
necessarily  be the case for 
observables sensitive to the detailed structure of 
multi-parton emission~\cite{Hautmann:2009zzb,Hautmann:2007gw}.  
For example, 
Fig.~\ref{fig:REFatlasm12}~\cite{Aad:2014qxa} shows ATLAS measurements of 
the di-jet invariant mass associated 
with $W$ production, 
compared with several   
  Monte Carlo calculations.  The 
comparison  with the results from 
the  NLO-matched  
 calculation  {\sc Blackhat}   
 + {\sc Sherpa}~\cite{Bern:2013gka}   suggests    
that effects beyond NLO + collinear   
shower may set in for high  invariant   
masses around and  above 500 GeV.   
In this region of masses 
a similar behavior is  observed 
in the comparison of 
experimental  measurements 
with   the 
{\sc Alpgen}~\cite{Mangano:2002ea}  Monte Carlo calculation.  
In   Fig.~\ref{fig:REFm12} we also plot the 
di-jet invariant mass distribution 
from the approach~\cite{Dooling:2014kia}.

\begin{figure}[htbp]
\begin{center}
\includegraphics[scale=0.3]{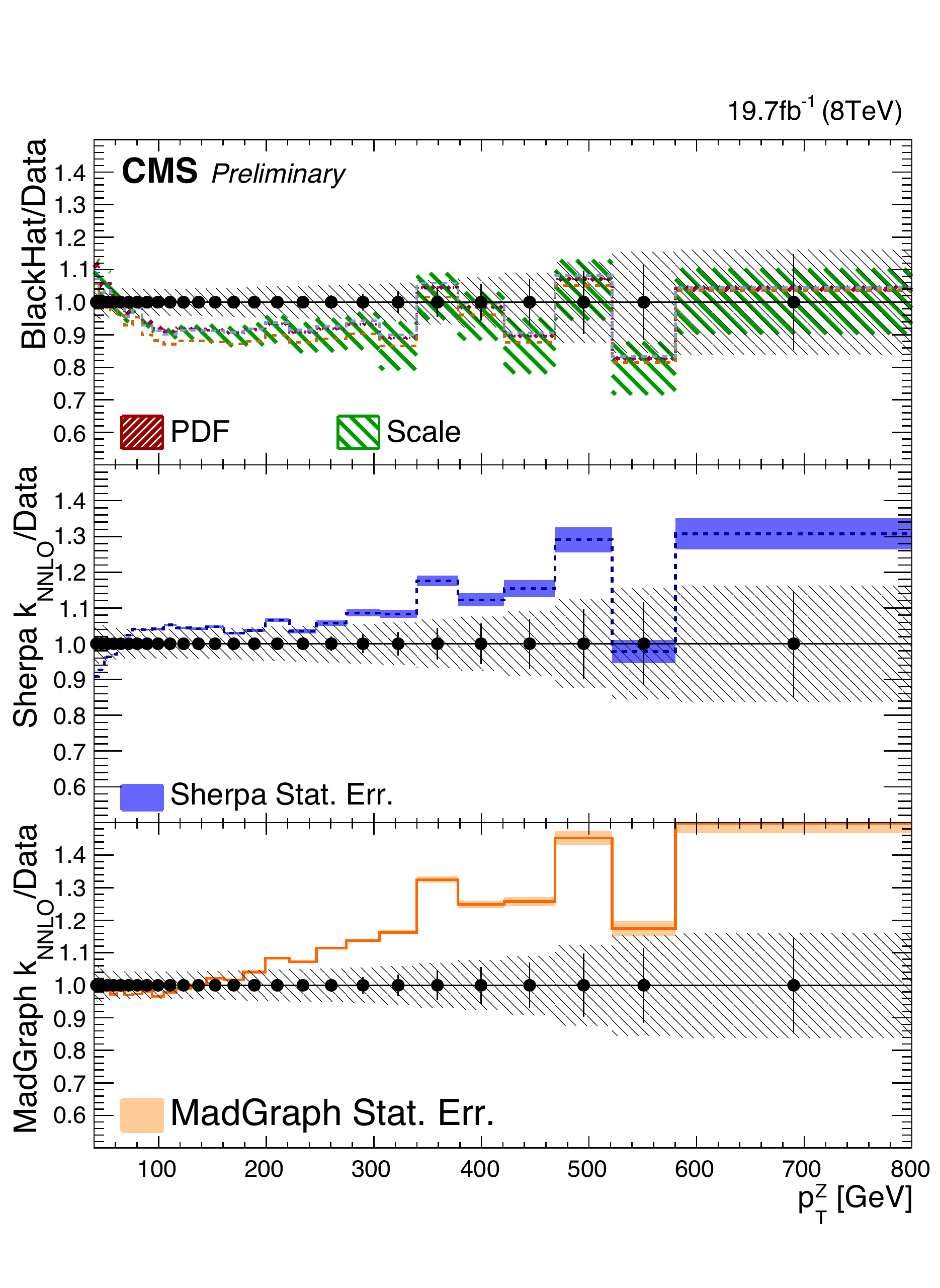}
\includegraphics[scale=0.3]{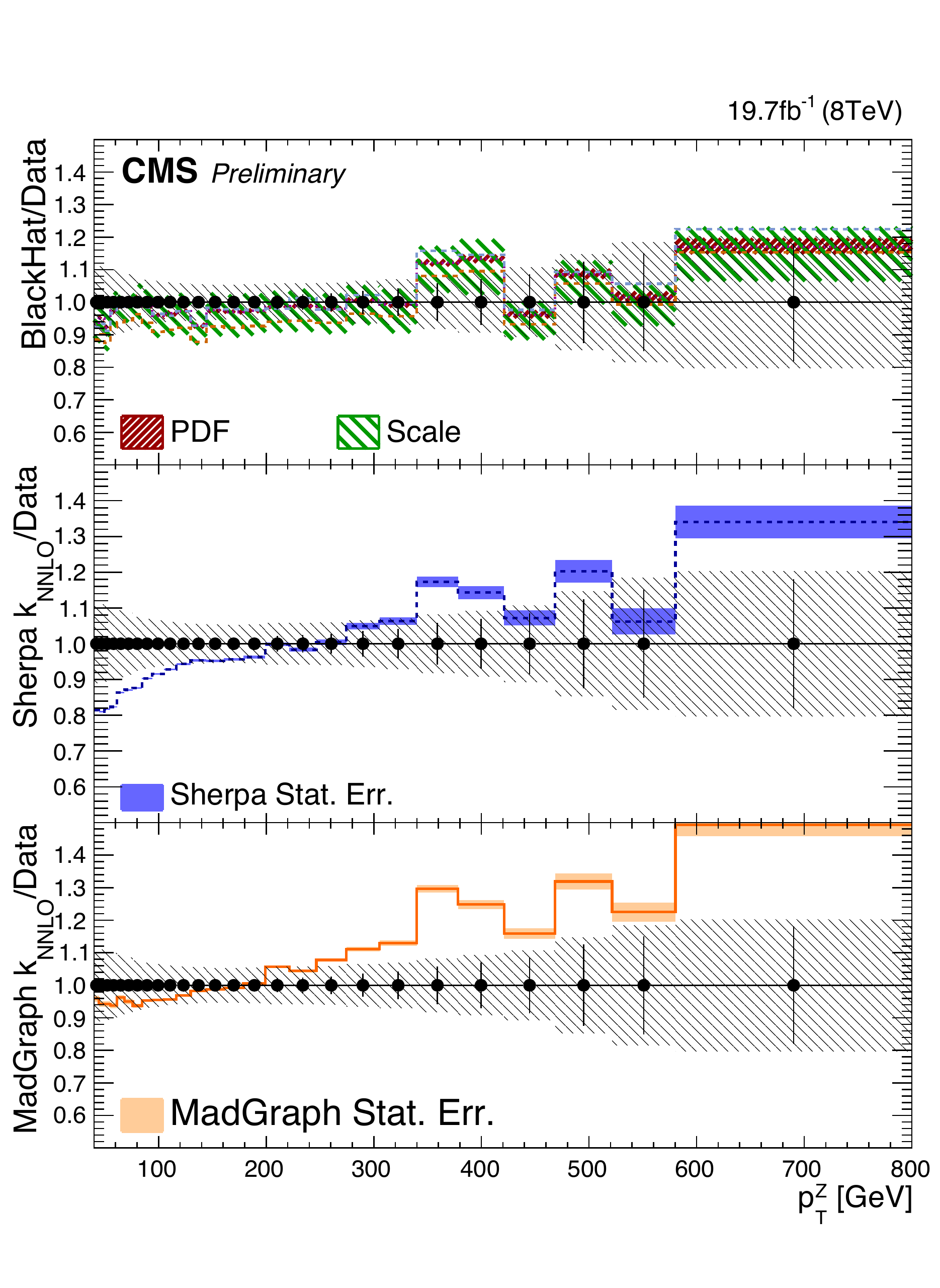}
\caption{\it   $Z$-boson 
transverse momentum 
measured~\protect\cite{Khachatryan:2015ira}
in 
$Z$ + 1 jet (left) and 
$Z$ + 2 jets (right) 
events, compared with   
Monte Carlo calculations.}
\label{fig:REFhighptZ}
\end{center}
\end{figure}

For the physics program at  Run II 
it is of much interest  to examine 
the region of very large  vector boson 
transverse momenta of 
order 1 TeV and higher.    
Fig.~\ref{fig:REFhighptZ}~\cite{Khachatryan:2015ira}   
shows    CMS   
measurements of the $Z$-boson 
$p_T$  
in events with  $Z$ + 1 jet and 
$Z$ + 2 jets  at Run I. 
At the highest $p_T$ one may see 
dynamics setting in beyond the level 
 of  {\sc Madgraph}~\cite{Alwall:2011uj} and 
{\sc Sherpa}~\cite{Gleisberg:2008ta} 
 multi-leg jet calculations matched with 
collinear 
showers, even  supplemented with 
an NNLO  k-factor.

\section{Theoretical developments and experimental prospects}

This section gives a brief overview  of  ongoing theoretical 
developments  and experimental prospects. 

\begin{description} 
\item
{\em Factorization and resummation  for $q_T \ll M $}. 
The factorization~\cite{Collins:1984kg,Collins:2011zzd}   
for Drell-Yan production at  
low  $q_T$ 
(along 
with  corresponding 
extensions to other processes,  including semi-inclusive 
DIS and  Higgs 
production)    
has been reobtained in 
  soft collinear effective 
theory   (SCET)  by  different 
approaches~(\cite{GarciaEchevarria:2011rb,Echevarria:2012js,Echevarria:2014rua,Idilbi:2010im,
Idilbi:2010tc},         \cite{Mantry:2009qz,Mantry:2010mk,Li:2011zp,Jain:2011iu,Stewart:2010qs},  
\cite{Becher:2010tm,Becher:2012yn,Becher:2011xn}).   
 The treatment of nonperturbative 
contributions  
to the TMD 
evolution equations~\cite{Collins:1981uw,Collins:1981va,Collins:2011zzd}
from the region 
of large 
transverse distances  ${ b}_T$
   differs   in each of these  various 
approaches and 
in the classic studies~\cite{Ladinsky:1993zn,Landry:2002ix,Konychev:2005iy,Guzzi:2013aja,Guzzi:2012jc,resbosweb},  
 and is currently the subject 
of  intense  investigations. 
Such  treatment is essential for predictions at 
$q_T \lsim 1 $ GeV  but its influence  may also  extend to the peak region.   
 It is found to be important, and with distinctive   
features compared to the Drell-Yan case, in semi-inclusive 
DIS~\cite{Boglione:2014oea,Boglione:2014qha}. 
See~\cite{Collins:2014jpa,D'Alesio:2014vja,Echevarria:2012pw,Boglione:2014oea,Boglione:2014qha,Bacchetta:in_prep,Aidala:2014hva,Schweitzer:2013iva,Schweitzer:2012dd,Su:2014wpa,Sun:2013hua,Prokudin:2015ysa}
for recent discussions of  
 nonperturbative contributions. 
The region of small 
transverse distances  ${ b}_T$, on the other hand,   is   
investigated via 
perturbative resummations 
to next-to-next-to-leading 
accuracy~\cite{Catani:2000vq,Bozzi:2005wk,Bozzi:2007pn,Catani:2010pd} 
and 
computations through 
two loops~\cite{Catani:2012qa,Catani:2011kr,Catani:2013tia,Gehrmann:2012ze,Gehrmann:2014yya}
of the perturbative 
coefficient functions 
controlling the expansion of the 
TMDs in terms of collinear 
pdfs.  All these aspects are 
relevant for the interpretation of 
the production spectra at low 
 transverse momenta  $q_T$, 
both in high-energy  Drell-Yan 
 experiments at LHC and Tevatron~\cite{Aad:2014xaa,Khachatryan:2015oaa,Aaltonen:2012fi,Abazov:2007ac,Affolder:1999jh,Abbott:1999yd,Abbott:1999wk} 
and in fixed-target 
experiments~\cite{Ito:1980ev,Moreno:1990sf,Antreasyan:1981uv}, including 
polarized Drell-Yan 
and semi-inclusive DIS~\cite{Adolph:2013stb,Airapetian:2012ki}.

\item
{\em Evolution of TMDs and fits to 
physical cross sections}. 
The above approaches 
to low-$q_T$ spectra 
which make  use of TMDs  
currently employ, in practice, 
either approximate analytic 
(or semi-analytic)   
solutions of the evolution 
equations~\cite{Collins:1981uw,Collins:1981va,Collins:2011zzd} 
or  perturbative  expansions of the TMDs in terms of collinear pdfs, or a 
combination of  both. A different 
proposal   has been put forward 
in~\cite{Hautmann:2014kza}    (\tmdlib),  
based  on    
global fits to 
 experimental data to obtain 
TMD  parton distributions at different 
evolution scales, and on  using  
 these  to make  predictions for physical quantities.  
This is  
similar in spirit (but   different 
in its realization) 
to what is done in the case of 
collinear parton distributions.  
Theoretical 
 predictions for physical cross 
sections which obey TMD factorization formulas could then be obtained by 
applying 
 these formulas,   using 
perturbatively calculable coefficients and appropriately evolved TMDs  determined 
from fits to experiment. 
In this approach, unlike  most 
current implementations of 
TMD formalisms, the  
nonperturbative dependence 
on longitudinal and 
transverse degrees of freedom 
is fully coupled,  
and can be entangled with the 
dependence on the 
evolution scale~\cite{Hautmann:2014kza}.  
For phenomenological 
applications this can be important  when for instance comparing theory 
with experimental measurements over a wide range in $x$ and  evolution scales.

\item
{\em Nonlinear evolution of the 
gluon TMD  and Wilson line  
correlators}.  
The conventional gauge-invariant operator definition of the 
gluon TMD~\cite{Collins:1981uw, Mulders:2000sh,Ji:2005nu,Meissner:2007rx,Sun:2011iw,Echevarria:2015uaa}  
is distinct  from the  
Weiszacker-Williams  operator 
definition~\cite{Dominguez:2011wm,Dominguez:2011gc,Mueller:2013wwa,Avsar:2012hj,Avsar:2011tz}  
   in  terms of  Wilson 
lines often used at 
 $x \ll 1 $ (see       also~\cite{Hautmann:2007cx,Hautmann:2006xc,Hautmann:2000pw} for discussion of  the  operator definitions).    Correspondingly, these gluon TMDs  obey different rapidity evolution  equations: in the moderate $x \sim 1$ region one has linear double-logarithmic equations, while 
in the $x \ll 1 $ domain the non-linear single-logarithmic Balitsky-Kovchegov equation applies~\cite{Balitsky:1995ub, Kovchegov:1999yj}. 
The relationship  
between these two 
regimes 
is examined 
in~\cite{Balitsky:2014wna,Balitsky:2015qba}, 
where it is clarified 
 that the non-linear small-$x$ evolution transforms into linear rapidity evolution for the conventional gluon TMD.  
 Refs.~\cite{Boussarie:2015dxa,Boussarie:2014lxa} 
consider applications to  
diffraction, and 
Refs.~\cite{Chachamis:2012mw,Chachamis:2013oga,Ducloue:2014koa,Ducloue:2013bva} 
to jets at large 
rapidity separations. 
Also, 
the  evaluation of the complex combinations of  Wilson lines entering the gluon TMD at small $x$ calls for the development of a dedicated methodology. Essential improvement in the understanding and computation of correlators with Wilson lines can be achieved by the eikonal exponentiation methods~\cite{Sterman:1981jc,Gatheral:1983cz,Frenkel:1984pz}, which enable the exact resummation of the diagrams presenting a given correlator as the exponent of series of the so-called 
web 
diagrams~\cite{Laenen:2008gt,Gardi:2010rn,Mitov:2010rp,Vladimirov:2014wga,Vladimirov:2015fea}.

\item
{\em TMDs and generalized loop space}. Renormalization 
 properties of 
Wilson line correlators control 
the evolution of TMDs~\cite{Collins:2011zzd,Cherednikov:2010uy,Cherednikov:2009wk,Cherednikov:2008ua,Cherednikov:2007tw}. 
In particular, 
the appearance of light-cone, or rapidity, divergences~\cite{Collins:2011zzd,Collins:1989bt} in higher-loop 
corrections  to the  gauge-invariant correlators 
 calls for a treatment of overlapping 
divergences, 
 which can be achieved by the  
introduction of a soft 
 subtraction factor~\cite{Hautmann:2007uw,Hautmann:2001yz,Collins:2000gd,Collins:1999dz,Chiu:2012ir,Fleming:2012kb}. 
The evolution of  the gauge-invariant path-dependent TMDs with 
the light-like cusped Wilson lines can also be associated with the geometric evolution in the generalized 
space~\cite{Cherednikov:2012yd,Cherednikov:2012ym,Cherednikov:2012qq}. 
The differential shape variations of the underlying contours to the Wilson loops are formulated in terms of the Fr\'echet derivative~\cite{Cherednikov:2014jva,Cherednikov:2014mua} and the equations of motion in the loop space  are dual to the energy and rapidity evolution of the TMDs having the same structure of the Wilson lines~\cite{Mertens:2013xga,Mertens:2014hma}.   
 
\item
{\em Non-universality and Wilson lines.}   
Operator definitions  of  parton distribution functions   
in terms of quark and gluon
fields involve nonlocal operator combinations. For collinear functions the
nonlocality is along the lightcone, for TMDs it is along the lightfront
involving also transverse separations. Unavoidably, therefore,   additional
gluonic fields minimally enter in the Wilson lines that are needed for
an unambiguous gauge invariant description. The fact that these Wilson lines
depend on the hard process brings in a calculable 
non-universality~\cite{Bomhof:2007xt}, which is a generalization of the sign
flip between T-odd TMDs in going from SIDIS to
DY~\cite{Brodsky:2002cx,Collins:2002kn}. Other examples where 
these effects appear are jet+jet or photon+jet final states in
hadroproduction~\cite{Boer:2011xd,Pisano:2013cya} as compared to Drell-Yan or $ ZZ $ 
production, and Higgs + jet final states  as compared to Higgs production
into colorless final states.  
Color entanglement  can lead to further 
sources of non-universality  affecting  both 
  TMD factorization~\cite{Rogers:2010dm}  and  
 collinear factorization~\cite{Catani:2011st,Forshaw:2012bi}. 
A program  is ongoing 
devoted to a careful analysis~\cite{Buffing:2012sz,Buffing:2013kca,Boer:2015kxa} 
of the 
possible operators that contribute to particular TMD structures followed by
the study of their evolution.

\item
{\em TMDs from exclusive evolution equations}.  
The gluonic   CCFM evolution  
equation~\cite{Ciafaloni:1987ur,Catani:1989sg,Marchesini:1994wr} 
is being extended along the lines proposed in~\cite{Hautmann:2014uua} 
to treat the coupled evolution of 
the flavor-singlet sea quark density and gluon density.   
This is  important for describing   exclusive components of 
high-multiplicity final states. 
In particular, the inclusion of the sea quark density  at TMD level  is one of the 
main elements needed to treat 
 Drell-Yan production  
across the whole range of central and forward 
rapidities~\cite{Hautmann:2012sh,Hautmann:2012pf,Banfi:2012du,Banfi:2011dm,Baranov:2014ewa,Baranov:2014tea,Lipatov:2014yna,Nefedov:2012cq} 
measured at the     LHC~\cite{Aad:2014xaa,Khachatryan:2015oaa,Aaij:2013nxa,Chatrchyan:2013oda,Aad:2013ysa}.
This approach is also being extended to include nonlinear evolution 
and saturation  effects~\cite{Kutak:2014wga,Deak:2015dpa,Deak:2015kra,Kotko:2015ura}  
and to incorporate methods for    automated computation of 
off-shell high-energy matrix   elements~\cite{vanHameren:2015bba,Bury:2015dla,vanHameren:2015wva,vanHameren:2014ala,vanHameren:2014lna,vanHameren:2015uia,Kotko:2014aba,vanHameren:2013csa}. 

\item
{\em Soft particle production and multi-parton interactions}.   As 
TMDs encode  nonperturbative transverse momentum dynamics in the proton,   
one may ask whether they are 
relevant  not only for factorization 
of hard processes but also for the  
understanding  of soft particle 
production and, in particular, of the 
multi-parton interactions which are 
found to be needed at low to 
moderate 
transverse momenta 
for Monte Carlo simulations to 
describe  
experimental data  
on underlying events, particle 
multiplicities and spectra. 
Double parton 
interactions~\cite{Paver:1982yp,Sjostrand:1987su}   including 
parton's transverse momentum dependence are  investigated
in~\cite{Kasemets:2012pr,Diehl:2014vaa,Diehl:2011yj,Maciula:2013kd,Maciula:2014oya,Maciula:2014xba,vanHameren:2014ava,Baranov:2015nma}. 
The role of parton's transverse momentum 
in the interpretation of 
 energy flow  measurements is 
discussed in~\cite{Hautmann:2013fla,Hautmann:2013yta,Hautmann:2012xa,Deak:2011ga}. 
Implications for diffraction are 
considered in~\cite{Pasechnik:2010zs,Kopeliovich:2007pq,Kopeliovich:2006bm}.  
TMD effects in 
multi-parton correlations 
may  be studied in  upcoming 
measurements  of 
 charged particle multiplicities and spectra and underlying event at the LHC 13 TeV run. 
\end{description}

Experimental prospects have been  
discussed  for 
identifying TMD effects  based on measurements of 
benchmark cross sections,  
 both at the LHC and at 
lower energy experiments.

\begin{description} 
\item
{\em Drell-Yan lepton pair production 
and Drell-Yan plus jets}.  
As discussed in the previous 
two sections, both the low-$q_T$ part of the spectrum and 
the high-$q_T$ part can be sensitive to TMD 
effects. Multi-differential 
measurements are  especially 
important as one can access  azimuthal 
correlations in the 
lepton + jet final states~\cite{Dooling:2014kia,Banfi:2012du} 
which constitute  distinctive  TMD predictions. 
Comparison of 
$ Z $ + jet final states at small transverse momentum imbalance~\cite{Boer:2011xd,Pisano:2013cya} 
 with  di-boson $ Z Z $  final states may shed light 
on color flow  patterns which are eventually 
responsible for factorization breaking phenomena 
in hard processes sensitive to 
 very low transverse momentum scales.

\item
{\em Higgs boson production and 
Higgs boson plus jets}. 
Similar measurements to the Drell-Yan case, including 
differential cross sections, 
are  relevant for 
  gluon TMDs and QCD studies of 
  polarized gluons and color 
correlations, once 
sufficient statistics is  reached. 
Measurements  of Higgs versus  
Drell-Yan at the same invariant mass 
may be used to  reduce the influence 
of 
pile-up in the high-luminosity LHC runs~\cite{Cipriano:2013ooa}.   
The  boson $q_T$ spectrum,   
final-state angular distributions 
and 
underlying event observables  
 probe  different  aspects of 
the Higgs coupling to 
gluons~\cite{Cipriano:2013ooa,Langenegger:2015lra}. 
 
\item
{\em Heavy flavor production}.   
Measurements of 
top quark 
pair production spectra 
can provide comparable 
information to the previous two 
cases but with additional complexity 
due to the presence of 
color charges in the final state. 
The associated 
initial-state / final-state 
color correlations 
at small $q_T$ 
could be studied to examine 
factorization-breaking 
contributions in the region of very 
small transverse momenta~\cite{Zhu:2012ts,Li:2013mia,Zhu:2013yxa,Catani:2014qha,Rogers:2010dm}, provided 
sufficient resolution can be 
reached. It will also be interesting 
to investigate kinematic effects 
of  longitudinal 
momentum reshuffling  in parton 
showers~\cite{Dooling:2012uw} at top quark  scales.  
Similar studies can be done at lower 
mass scales with bottom and charm 
quarks.  

\item
{\em Quarkonium production}. 
Despite the complexity of the bound state,   production of 
$c {\bar c} $ and $ b {\bar b}$ 
quarkonia is a useful probe of TMD gluon effects at low mass scales. 
Phenomenological studies 
are carried out  in~\cite{Baranov:2012fb,Baranov:2011ib,Baranov:2007ay,Lansberg:2013qka,Lansberg:2011hi,Dunnen:2014eta,Pisano:2014yea,Boer:2012bt,Ma:2012hh,Ma:2014oha}.   
Many features of these processes 
have been investigated 
experimentally at the LHC 
Run I~\cite{Chatrchyan:2013yna,YORK:2013ixa,Khachatryan:2010yr,Zheng:2012ema,Abelev:2012rz,PorteboeufHoussais:2012gn,Porteboeuf:2010dw}. 
Measurements of the spectra and especially of the polarization  for 
 $ J/ \psi$, $\Upsilon$ and all  quarkonium states  at Run II 
will be  particularly interesting for studying 
 polarized gluon effects. 
Color-singlet transitions may have reduced 
sensitivity to factorization-breaking 
effects~\cite{Dunnen:2014eta,Boer:2012bt,Ma:2012hh}. 
Quarkonium  measurements are 
further proposed at 
 fixed-target 
experiments~\cite{Lansberg:2012kf,Lansberg:2014myg,Massacrier:2015nsm} 
and electron-ion collider~\cite{Boer:2010zf,Pisano:2013cya}. 

\end{description}

\section{Working with TMDs: 
fits and parameterizations}
\label{s:parametrizations}
  
The polarization dependent 
and transverse momentum 
dependent 
proton's parton densities, in the 
notation of~\cite{Mulders:1995dh,Buffing:2012sz,Boer:1997nt,Buffing:2013kca,Boer:2015kxa},    
are  given  in the Tables~I and~II in Sec.~II. 
This scheme can be extended to spin-1  targets~\cite{Bacchetta:2000jk}.

Most existing 
fits and parameterizations 
 of these 
distributions  may be grouped  into 
  categories which 
broadly correspond to three main areas 
of  TMDs applications discussed 
in Sec.~II: 
\begin{itemize}
\item Fits to vector boson $q_T$ experimental  data 
in unpolarized Drell-Yan 
production~\cite{Landry:2002ix,Konychev:2005iy,Guzzi:2013aja,resbosweb,Qiu:2000hf,Qiu:2000ga,D'Alesio:2014vja,Melis:2015upa,Echevarria:2012pw,Su:2014wpa,Sun:2013hua}  
 based on  the low-$q_T$ TMD 
factorization~\cite{Collins:1984kg,Collins:2011zzd},  
in some cases including extension to 
semi-inclusive DIS  data~\cite{Adolph:2013stb,Airapetian:2012ki,Alexa:2013vkv}. 
\item Fits to DIS structure function 
data~\cite{Hautmann:2013tba,Jung:2002wn,Hansson:2003xz,Jung:2012hy,Kutak:2012rf,Kimber:2001sc,Kwiecinski:1997ee,Blumlein:1995eu,Blumlein:1995mi,Luszczak:2013rxa,Kowalski:2010ue,Kowalski:2011zza,Alekhin:2014irh,Levin:2013tca,Hentschinski:2012kr,Grinyuk:2013tt,Grinyuk:2012mc,Lipatov:2013yra,Albacete:2010sy,Iancu:2015joa,Rezaeian:2013tka,Kuokkanen:2011je} 
  based on the high-energy TMD 
factorization~\cite{Catani:1990eg,Catani:1990xk}
or on  other  approaches (e.g.  saturation formalism)   to high-energy DIS, in some cases including the 
precision  measurements~\cite{Aaron:2009aa,Abramowicz:1900rp} 
and   TMD pdf 
   uncertainties~\cite{Hautmann:2013tba,Alekhin:2014irh}. 
\item Fits to spin and azimuthal asymmetries 
 data from low-energy experiments  either  based on 
parton model~\cite{Schweitzer:2010tt,Meissner:2009ww,Meissner:2008ay,Avakian:2010br,Avakian:2009jt,Efremov:2009ze,Efremov:2010mt,Efremov:2004tp,Collins:2005ie,Anselmino:2005nn,Anselmino:2006yc,Anselmino:2007fs,Avakian:2007mv,Boffi:2009sh,Arnold:2008ap,Barone:2009hw,Vogelsang:2005cs,Bianconi:2005yj,Bacchetta:2011gx,Signori:2013mda,Signori:2013gra,Radici:2015mwa,Barone:2015ksa,Godbole:2012bx,Godbole:2013bca,Godbole:2014tha,Mao:2014fma,Zhang:2008ez,Zhang:2008nu,Boer:2010zf,Boer:2011xd,Pisano:2013cya,Qiu:2011ai} 
or including QCD evolution~\cite{Kang:2014zza,Kang:2015msa,Echevarria:2014xaa,Sun:2013dya,Bacchetta:2015spa,Boer:2015vga,Aidala:2014hva,Aybat:2011zv,Aybat:2011ta,Ceccopieri:2014qha,Ceccopieri:2007ek,Ceccopieri:2005zz,Anselmino:2012aa,Anselmino:2013rya,Anselmino:2013vqa,Anselmino:2014eza,Anselmino:2014pea}. 
\end{itemize} 

For precision phenomenology it will 
be essential that  results of 
fits and parameterizations are given 
in a  portable  form as 
a determination of 
 TMD pdfs  over a given kinematic 
range, appropriate to the theoretical 
method and experimental data used. 
A first step in this direction has been 
taken in~\cite{Hautmann:2014kza}. 
The main point  is that if  results of 
fits to experimental data 
are used to provide 
 TMD pdfs at different evolution scales, 
theoretical predictions for physical cross sections could then be 
obtained by using these pdfs in 
factorization formulas (or, eventually, 
in Monte Carlo event generators 
implementing these formulas). 
In~\cite{Hautmann:2014kza}
 a library has been initiated, 
\tmdlib, 
 in order to unify and simplify the access of TMDs, 
along with a plotting tool, 
\tmdplotter, for easier comparisons. 
Commonly used pdf sets  are implemented  in \tmdlib, with  
the goal to provide a  library of all available TMDs. In \tmdlib\  pdfs are accessible in an easily callable way within the range of their applicability. The pdfs currently included 
range from TMD gluon densities obtained 
from fits to small-$x$ DIS data  
based on high-energy factorization, to TMD gluon densities from fits based 
on  saturation approaches, to  TMD quark densities from parton-model 
fits to low-energy fixed-target data 
at large $x$ and small $k_T$. 
 TMD fragmentation functions are not yet implemented, but are foreseen for the future.

\begin{figure}[htbp]
\begin{center}
\includegraphics[scale=0.5]{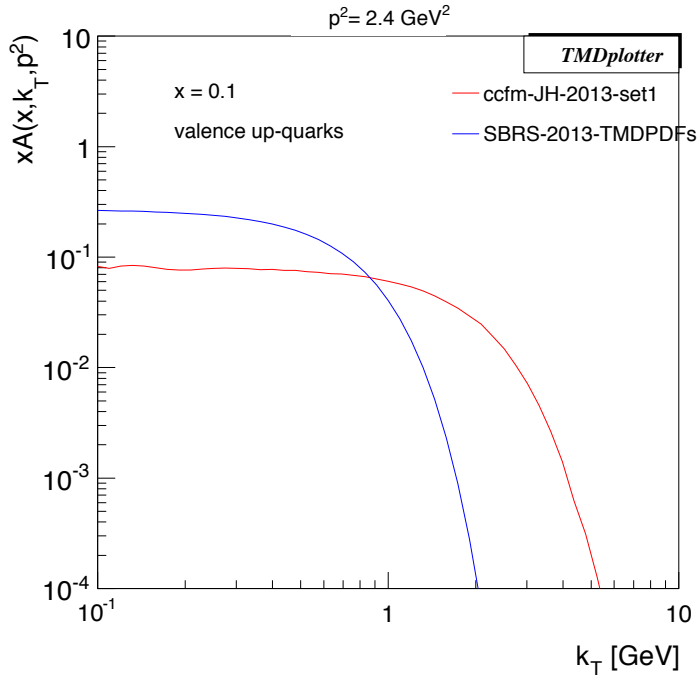}
\caption{\it  Valence quark distributions as a function of 
transverse momentum~\protect\cite{Hautmann:2014kza}  from  
the fits~\protect\cite{Signori:2013mda,Hautmann:2013tba}.}
\label{fig:tmdplotter}
\end{center}
\end{figure}

An example  from 
\tmdlib\ is shown in 
Fig.~\ref{fig:tmdplotter}, 
plotting the transverse momentum 
dependence of  valence quark 
distributions, at fixed values 
 of $x$ and renormalization   scale 
$p^2$, obtained 
from the 
fits~\cite{Signori:2013mda,Hautmann:2013tba}. 

\begin{figure}[htbp]
\begin{center}
\includegraphics[scale=0.5]{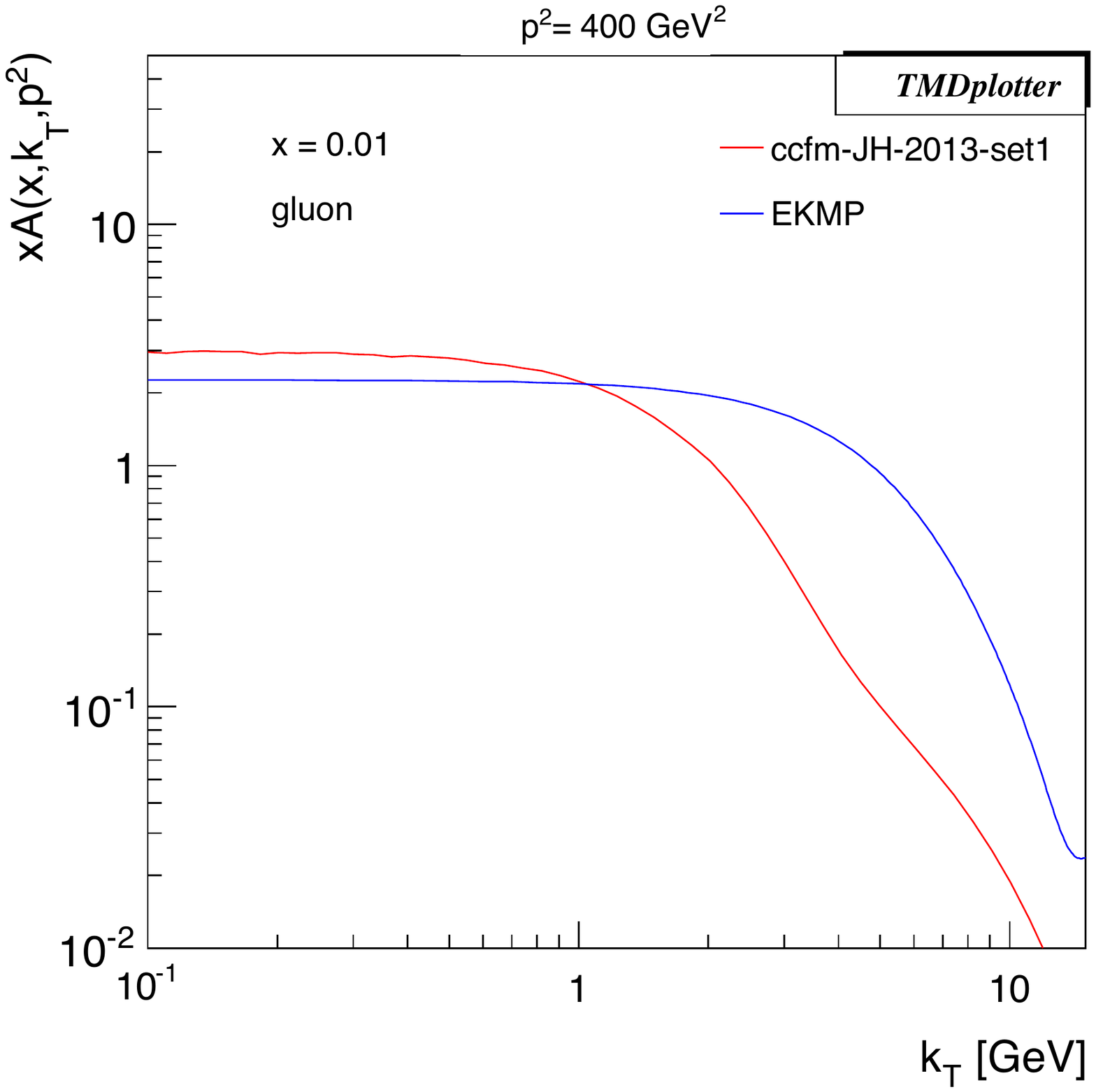}
\caption{\it 
Gluon distributions from~\protect\cite{Echevarria:2015uaa,Hautmann:2013tba}
as a function of 
transverse momentum~\protect\cite{Hautmann:2014kza}.}
\label{fig:tmdplotter-glue}
\end{center}
\end{figure}

In Fig.~\ref{fig:tmdplotter-glue} we  show   results for  gluon distributions~\cite{Echevarria:2015uaa,Hautmann:2013tba}. 
The nonperturbative parameters of the distribution represented by the red curve  are obtained  from  the fit~\cite{Hautmann:2013tba} to   DIS experimental  data~\cite{Aaron:2009aa,Abramowicz:1900rp} 
while  the  distribution represented by the blue curve is not fitted but based on the   model~\cite{Echevarria:2015uaa}.

\section{Working with 
TMDs: Monte-Carlo Generators and Tools}

Inclusive or semi-inclusive hard 
cross sections can be calculated 
by convoluting  parton density 
and decay 
 functions with   partonic cross sections.  For a detailed description 
 of the 
exclusive structure of the 
 final states, on the 
other hand, event generators  
including  parton showers 
  and full hadronization are  
  required.

In the collinear case,  cross sections are computed with on-shell  initial partons. For many processes, higher order calculations exist, and many of these are implemented in Monte Carlo (MC) simulation tools like 
{\sc  Powheg}~\cite{Nason:2004rx,Frixione:2007vw}, {\sc Mc@nlo}~\cite{Frixione:2002ik}, a{\sc  Mc@nlo}~\cite{Alwall:2014hca}, which combine next-to-leading order partonic calculations with parton showers and hadronization. These simulations all need  a reshuffling of kinematic variables, after the parton shower is generated, in order to satisfy 
energy-momentum conservation,  which  
 can lead to significant kinematic shifts in the longitudinal momentum fraction $x$~\cite{Dooling:2012uw}. This is because    transverse momentum is   generated by the initial-state parton shower, which is not available  when the hard scattering is computed. 
In certain phase space regions, 
these longitudinal shifts 
can affect the accuracy of the 
calculations significantly. 
Using TMDs, this kinematic reshuffling can be avoided from the beginning provided the TMDs include transverse momenta generated by perturbative QCD evolution, which 
in turn can be evaluated 
according to different approximation 
schemes such as those in~\cite{\DGLAP},~\cite{\BFKL},~\cite{Ciafaloni:1987ur,Catani:1989sg,Marchesini:1994wr}.

If  a Monte Carlo method is used to solve the TMD  evolution equation, 
a further advantage is 
 that  the solution of the evolution equation can be directly matched to the simulation of parton showers: the kinematic distributions are the same, whether they come from a solution of the evolution equation or from a simulation of the parton shower~\cite{Jung:2000hk,Tanaka:2014yxa,Hautmann:2014uua}.

While a general purpose Monte Carlo 
at TMD level does not yet exist, examples of such 
algorithms~\cite{Schnell:2015gaa,Avakian:2015faa,Avakian:2015vha} have been presented for specific cases. We list 
a few examples below.

\begin{itemize}
\item{MC event generators with parton shower and hadronization}

$\blacktriangleright$  \cascade\  \cite{Jung:2000hk,Jung:2001hx,Jung:2010si,cascadeweb}
is a full hadron level Monte Carlo event generator using TMDs, originally developed for small $x$ processes in $ep$, now extended to cover medium and large $x$ and $pp$ processes. Initial state parton showers are treated according to the CCFM formalism, final state parton shower and hadronization is performed by the Lund package \pythia\ \cite{Sjostrand:2006za}. Parton polarizations are included according to the high-energy factorization~\cite{Catani:1990eg}. Proton polarizations are not yet included.

$\blacktriangleright$ \pythia\ \cite{Sjostrand:2006za}. 
With the initial and final state parton showers simulated in \pythia ,    one may argue that several elements of TMD physics are effectively included. \pythia\ can be used to mimic spin-dependent cross sections by reshuffling events (assigning polarization states) \cite{Schnell:2015gaa} according to a given cross-section model. This is especially useful when event topologies are needed (e.g., to simulate the interplay of track correlations with detector performance), or where no explicit physics model is yet available to be employed in dedicated MC generators. 


$\blacktriangleright$ {\sc mPythia}  and {\sc mLepto}  
are  based on {\sc Lepto} \cite{Ingelman:1996mq} and \pythia\ \cite{Sjostrand:2006za} with a modification of the hard process  \cite{Schnell:2015gaa} to treat the azimuthal angle of the scattered (light) quark and via momentum conservation of the target remnant according to parameterizations of the Sivers function. While limited to the rather specific case of the Sivers effect it can make use of the hadronization embodied in {\sc Jetset} \cite{Sjostrand:1985ys,Sjostrand:1993yb,Sjostrand:1995iq}. 

\item{MC event generators at parton level with fragmentation functions}

$\blacktriangleright$ {\sc LxJet}  (see \cite{LxJet}) 
 is devoted to a calculation of jet cross sections at small $x$ in hadron-hadron collisions. It can be also viewed as an event generator as it allows one to generate unweighted events. It uses high-energy factorization~\cite{Catani:1990eg}. 

$\blacktriangleright$ {\sc GMC-Trans}  (see \cite{Schnell:2015gaa})  
is a MC generator, developed by the HERMES Collaboration, applying the parton-model expression of the one-hadron semi-inclusive DIS cross section using several models/parametrization for various leading-twist TMD PDFs and FFs. Pion and charged-kaon production is simulated, both for proton and neutron targets (or combinations thereof) without including nuclear effects. 
An analytic expression for the semi-inclusive DIS cross section was implemented based on the widely used Gaussian ansatz of the transverse-momentum dependences.

$\blacktriangleright$ {\sc TMDGen} (see  \cite{Schnell:2015gaa})   
is an extended version of {\sc GMC-Trans} entirely  written  in  C++ focusing mainly on di-hadron production in semi-inclusive DIS. Advances in computation power allowed for other than the Gaussian ansatz of the transverse-momentum dependences by employing numeric integration algorithms. It thus allowed the usage of the 
spectator model~\cite{Bacchetta:2008af}  for various TMD PDFs and FFs.

$\blacktriangleright$ {\sc Clas} (see  \cite{Schnell:2015gaa})   
uses a similar approach as {\sc GMC-Trans}, though restricted to the unpolarized sector and to longitudinal double-spin asymmetries. It uses the fully differential single-hadron DIS cross section to simulate semi-inclusive DIS events. The transverse momentum dependence can be Gaussian, but also light-cone quark-model inspired dependence has been implemented. 

\item{Semi-analytical calculations of semi-inclusive processes}

$\blacktriangleright$ {\sc Resbos} \cite{Ladinsky:1993zn,Landry:2002ix,resbosweb}   
is a package to calculate analytically resummed distributions of inclusive and semi-inclusive observables. The $q_T$ resummation in {\sc Resbos} and  parton showering methods of Monte Carlo event generators are complementary. Both are based  on all-order resummation  using Sudakov form factors.
 {\sc Resbos} allows the user to calculate resummed distributions 
of the Higgs/vector bosons and their decay products up to NNLL. 
It follows a prescription for matching the 
resummed contribution  onto the fixed-order result and implements 
a parameterization of non-perturbative effects at small $q_T$ 
in terms of TMD PDFs.    

$\blacktriangleright$  {\ttfamily HqT} and 
{\ttfamily DYqT}~\cite{Bozzi:2005wk,Bozzi:2010xn} 
are numerical programs which implement the 
analytical $q_T$ resummation  
 formalism~\cite{Catani:2000vq,Bozzi:2005wk,Bozzi:2007pn,Catani:2010pd}
 to compute, respectively, the $q_T$ spectrum of the Standard Model Higgs and 
Drell-Yan lepton pair (via vector boson production) in hadronic collisions. 
The resummed results are  matched to the fixed order 
calculation valid at high $q_T$. The program can be used up to  
NNLL+NLO, with the resummed part evaluated at NNLL, the fixed order
evaluated at NLO (Higgs/vector bosons plus one or two partons) 
and with the normalization fixed to the total NNLO cross section. 

$\blacktriangleright$ {\sc HRes} and 
{\sc DYRes}~\cite{dyres:in_prep,deFlorian:2012mx} 
are numerical programs which 
extend the calculations in  {\ttfamily HqT}/{\ttfamily DYqT} 
by retaining the full kinematics  of the Higgs/vector bosons and of 
its decay products. The programs 
implement  $q_T$  resummation up to NNLL+NNLO and allow the 
user to apply arbitrary cuts on final states and to plot the corresponding
distributions in form of bin histograms. 
\end{itemize}

\section{Conclusions}

We  studied two sets of  examples of 
 multi-scale problems in hadronic collisions   which require 
QCD factorization theorems beyond the collinear approximation and 
call for the use of TMD parton distributions. 
In one set of examples,   the transverse momentum scale is small compared 
to the hard process scale; in the other, the transverse momentum 
is of the order of the hard scale but this is much smaller than the 
total energy of the scattering. In both cases, factorization 
theorems  in terms of TMD parton distributions are necessary in 
order both  to resum  logarithmically-enhanced perturbative 
corrections to all loops and to properly 
take into account nonperturbative hadron structure effects. 

 These  
 multi-scale regimes  are  relevant 
 to LHC phenomenology. An example is  
the low-$q_T$ region of transverse momentum spectra 
for vector bosons, Higgs bosons, heavy flavor pairs at the LHC. 
 Another example is the  production of multi-jets 
 associated with heavy bosons and heavy flavors at large 
 jet  masses. 
 Further examples include any final state produced 
by  events at small longitudinal momentum fraction $x$, such 
as final states boosted to high  rapidities.   
Besides the LHC, TMD dynamics is central to spin 
physics in current low-energy  experiments and to the planning 
of future polarized collider and fixed-target experiments.  

As the field moves towards  the  stage of precision studies, 
appropriate phenomenological tools will be  needed. This includes 
tools for  Monte Carlo event  simulations,  which require  
  parton shower evolution algorithms   
and  determinations  of TMD parton distributions from experimental 
data. First steps toward a  new  program 
of portable and accessible TMD pdfs  were illustrated with 
explicit examples in this  report.

\vskip 0.8 cm 

\noindent 
{\bf Acknowledgments}.  We are grateful to  DESY, NIKHEF and 
 Antwerp University   for 
financial  support to   the 2014 Workshops on  ``Resummation, Evolution,  Factorization" (REF 2014).   
IOC  acknowledges  support from   the Belgian Federal Science Policy Office.
The work of MGE and AS is part of the program of the Stichting voor Fundamenteel Onderzoek der Materie (FOM), which is financially supported by the Nederlandse Organisatie voor Wetenschappelijk Onderzoek (NWO). 
The work of PJM is part of the EU Ideas programme QWORK (Contract No. 320389).  
FH thanks the University of Hamburg and DESY for hospitality. 
The work of FH  is  supported in part  by    the DFG SFB 676 programme  Particles, Strings and the Early Universe. 


\end{document}